\documentclass[10pt]{article}

\usepackage[utf8]{inputenc}
\usepackage[a4paper, portrait, margin=1in]{geometry}
\usepackage{comment}
\usepackage{float,placeins}
\usepackage[dvipsnames]{xcolor}
\usepackage{bbm}
\usepackage{makecell}
\usepackage{multirow}
\usepackage{multicol}
\usepackage{subcaption}
\usepackage{booktabs}

\usepackage{amsmath}
\DeclareMathOperator*{\argmax}{arg\,max}

\usepackage[backend=biber, maxcitenames = 2, mincitenames=1,
citestyle=authoryear, giveninits=true, isbn=false, doi=true, url = true,
maxbibnames=99, sorting=none, style=numeric,
uniquelist=false,uniquename=init,sortcites]{biblatex}

\bibliography{MF, mirluPURC, MadsBibAddon}

\newcommand*{\tabindent}{ \hspace{3mm}}

\newcommand{\tr}{{\mbox{\scriptsize T}}}
\definecolor{mygray}{gray}{0.6}

\usepackage{authblk}
\usepackage{amsmath,amssymb}
\usepackage{bm}
\usepackage{graphicx}
\usepackage[colorlinks=true, allcolors=blue]{hyperref}
\usepackage{cleveref}

\usepackage{color}
\def\tr{\color{red}}

\usepackage{appendix}

\usepackage{float}

\def\tr{\color{red}}

\usepackage{caption}
\usepackage{subcaption}
\usepackage{bm}
\usepackage{xr}

\newlength{\tvl}
\newcommand{\tabtrip}[3]{$\,$\makebox[\tvl][r]{#1} $\;$[\textit{\makebox[\tvl][r]{#2}, \makebox[\tvl][r]{#3}}]}
\newcommand{\tabsingle}[1]{$\,$\makebox[\tvl][r]{#1} \phantom{$\;$[}\textit{\makebox[\tvl][r]{}\phantom{,} \makebox[\tvl][r]{}}\phantom{]}}

\title{Bikeability and the induced demand for cycling}

\author[a]{Mogens Fosgerau}
\author[b]{Miros\l{}awa \L{}ukawska} 
\author[b]{Mads Paulsen}
\author[b]{Thomas Kjær Rasmussen}

\affil[a]{University of Copenhagen,  E-mail: mogens.fosgerau@econ.ku.dk}
\affil[b]{Technical University of Denmark}

\begin{document}


\maketitle

\begin{abstract}
    To what extent is the volume of urban bicycle traffic affected by the provision of bicycle infrastructure? In this study, we exploit a large dataset of observed bicycle trajectories in combination with a fine-grained representation of the Copenhagen bicycle-relevant network. We apply a novel model for bicyclists' choice of route from origin to destination that takes the complete network into account. This enables us to determine bicyclists' preferences for a range of infrastructure and land-use types. We use the estimated preferences to compute a subjective cost of bicycle travel, which we correlate with the number of bicycle trips across a large number of origin-destination pairs.  Simulations suggest that the extensive Copenhagen bicycle lane network has caused the number of bicycle trips and the bicycle kilometers traveled to increase by 60\% and 90\%, respectively, compared with a counterfactual without the bicycle lane network. This translates into an annual benefit of €0.4M  per km of bicycle lane owing to changes in subjective travel cost, health, and accidents. Our results thus strongly support the provision of bicycle infrastructure. 

\end{abstract}
\paragraph*{Author contributions}{Author contributions: M.F designed research; M.P performed research;  M.F., M.\L{}., M.P., and T.K.R. analyzed data; M.F., M.\L{}., M.P., and T.K.R. wrote the paper.}
\section*{Significance statement}{Promotion of bicycle use has received considerable global attention. In addition to reducing the impact of urban transportation on climate, it will help to improve public health, and reduce traffic congestion, noise, and air pollution. Provision of bicycle-friendly infrastructure is a primary means to achieving this. Using a large dataset of observed bicycle trip trajectories and fine-grained network data covering the city of Copenhagen, Denmark, this study finds a large effect of infrastructure provision on the volume of bicycle traffic.}

\begin{figure}[ht]
\begin{subfigure}[t]{0.49 \textwidth}
\centering
          \includegraphics[width = \columnwidth]{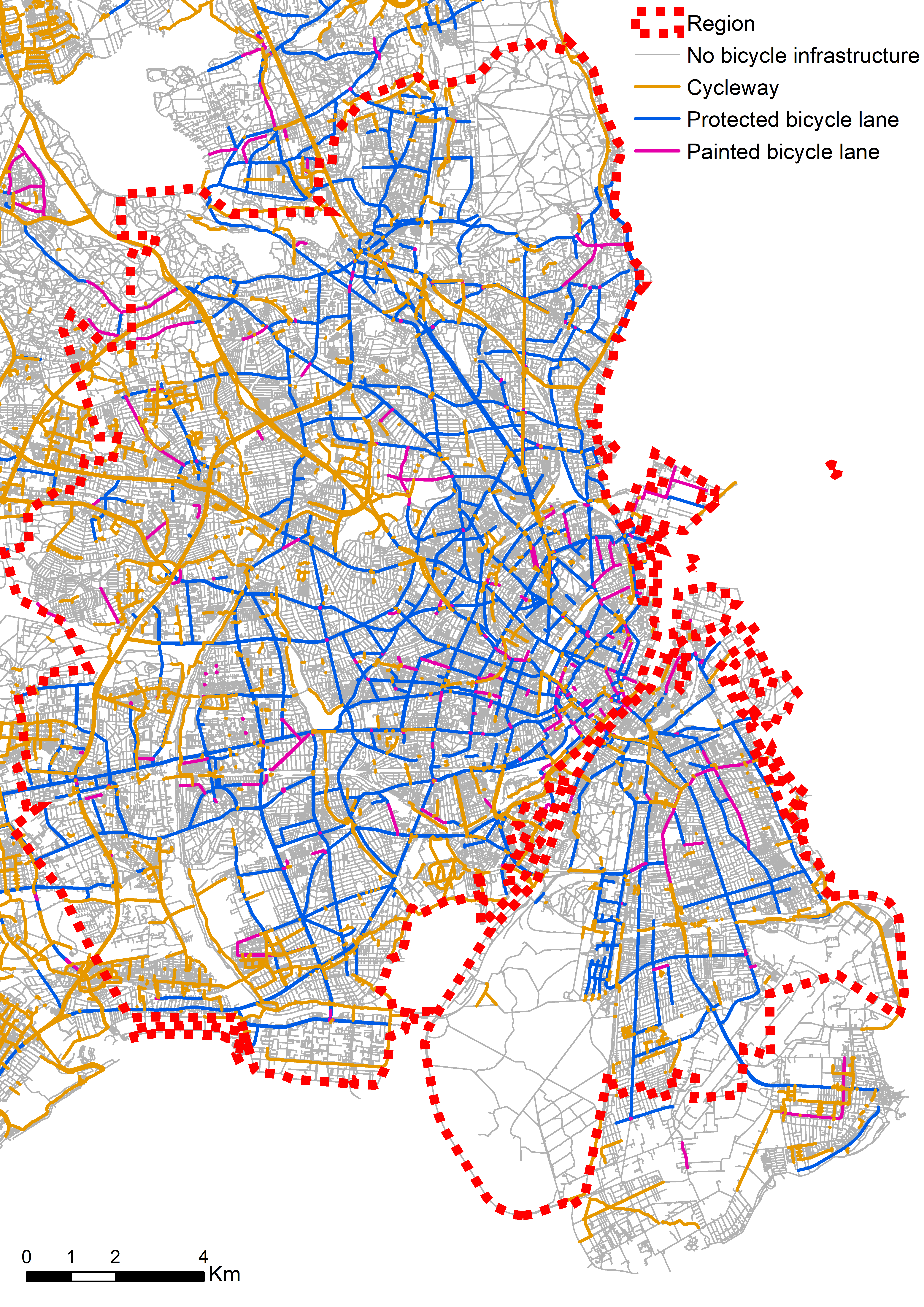}
 \caption{Bicycle-relevant infrastructure in Greater Copenhagen. The gray links indicate parts where it is possible to bicycle albeit with no dedicated bicycle infrastructure available.}
     \label{fig:bicyclenetwork}
\end{subfigure}
\hfill
\begin{subfigure}[t]{0.49 \textwidth}
\centering
 \includegraphics[width = \columnwidth]{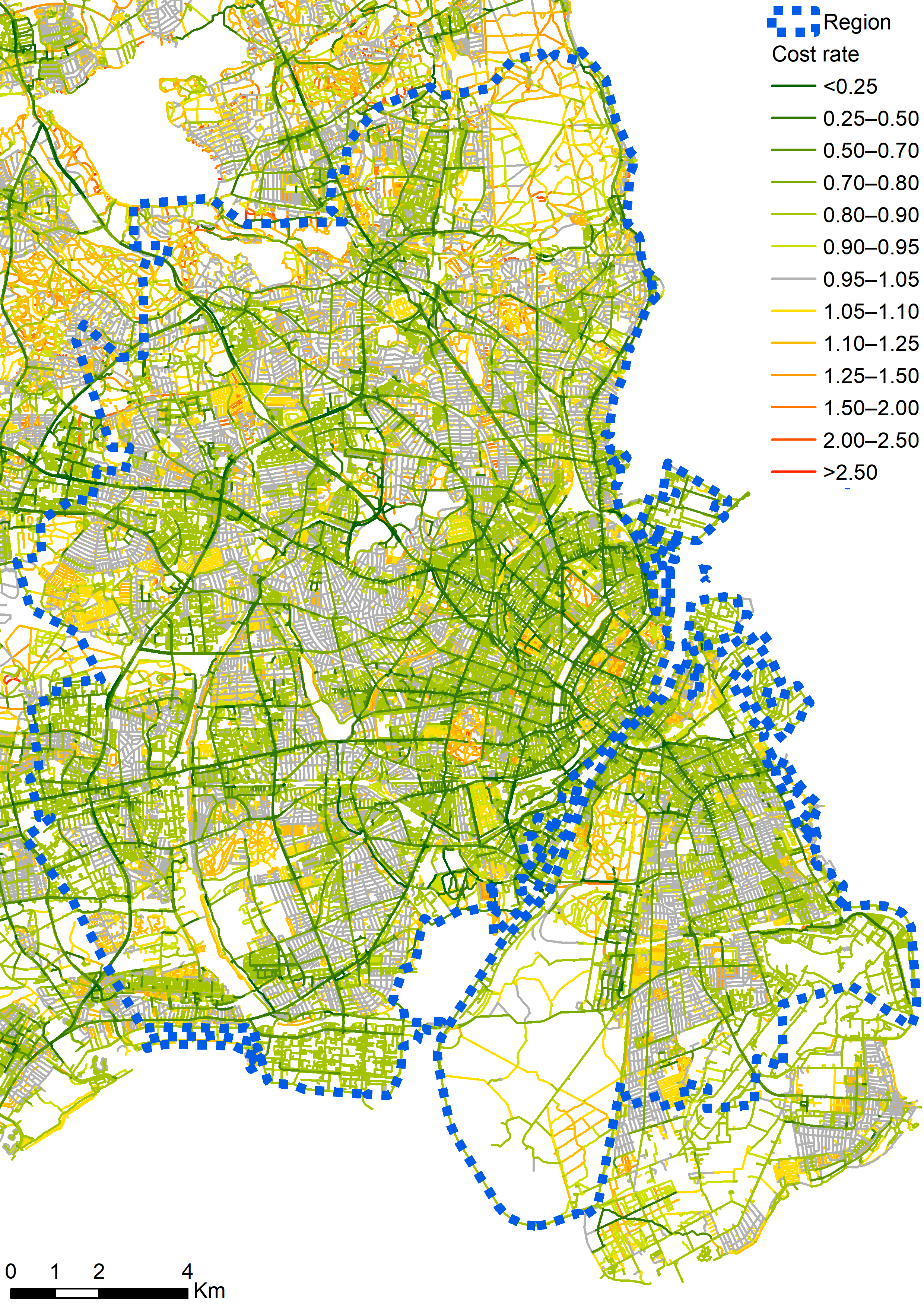}
 \caption{Estimated link cost rates; gray links mark the reference case: residential roads with no bicycle infrastructure in low-rise urban areas, scaled to a value of 1. Shades of green correspond to increasingly more attractive links, and shades of orange and red correspond to increasingly less attractive links.}
    \label{fig:NetworkCostRates}
\end{subfigure}
\caption{Maps representing the network used in the case study.} \label{fig:NetworkAndUnitCost}
\end{figure}

Copenhagen has extensive bicycle infrastructure and a high level of bicycle usage for everyday urban travel \parencite{CSC:18}. We have a dataset of unprecedented size available, comprising 218,489 bicycle trajectories in Copenhagen obtained from users of Hövding airbag helmets \parencite{Hovdingdata}. Matching these trajectories with very detailed network information (see Figure \ref{fig:bicyclenetwork}) allows us to track the observed bicycle route choices across a range of infrastructure and land use types.  

To make inference regarding the factors influencing bicyclists' choice of routes, we must compare the observed chosen routes to the possible alternatives. However, the number of possible routes between two points in a large network is extremely large and impossible to enumerate. To overcome this, we exploit the recently proposed perturbed utility route choice model \parencite{Fosgerau2021a}, which allows the entire network to be taken into account while being computationally feasible. 

The estimated route choice model indicates very substantial variation in the subjective cost of using various infrastructure types. The subjective cost per meter traveled on the most attractive infrastructure type, a cycleway, designated as a cycle superhighway, and located in a green area, is seven times lower than the subjective cost of cycling on a residential road.  

Bicycle infrastructure can therefore considerably affect the subjective cost of cycling and thereby the volume of bicycle use. We relate the number of trips in each origin-destination (OD) pair to the subjective cost from the route choice model. Employing a variant of the gravity model \textcite{Wilson1967}, we find a clear relationship whereby lower subjective cost is associated with a higher number of bicycle trips. We use this model to simulate a range of counterfactual scenarios, exploring the impact of the bicycle network on the volume of bicycle use.

\section{Results}

\subsection{Bicycle route choice}

We employ the perturbed utility route choice model \textcite{Fosgerau2021a}, which is a perturbed utility model \textcite{McFadden2012,Fudenberg2015,Allen2019}, adapted to describe the route choice through a network. For each OD pair, the predicted behaviour of bicyclists is a vector $x\in \mathbb{R}^{|\mathcal{E}|}_+$ that represents the distribution of flow across the network links $e\in\mathcal{E}$. The model assumes that the observed flow $x$ minimizes a convex cost function under the constraint that the flow $x$ is physically consistent with a flow of mass one through the network from origin to destination. The cost does not involve monetary elements but is a subjective cost that represents the bicyclists' preferences for various infrastructure types. The cost function has the form
\begin{equation}
C(x) = \sum_{e\in\mathcal{E}} l_e  \left( c_e  x_e + F(x_e) \right), 
 \label{eq:U}
\end{equation}
where $l_e$ are the link lengths, $x_e$ are the link flows, and link cost rates are specified as  $c_e=z_e' \beta$, where $z_e$ is a vector of link characteristics and $\beta$ is a vector of the parameters to be estimated. The perturbation function $F(\cdot):\mathbb{R}_+ \rightarrow \mathbb{R}$ is defined as $F(x_e) = (1+x_e) \ln{(1+x_e)} -x_e$, which is a convex function with $F(0)=F'(0)=0$. The perturbation term provides incentive to distribute flow on more than one route, while allowing the cost-minimizing flow to be zero on most links. 

The materials and methods section explains the data and estimation methods, and further details are provided in the SI Appendix. The full specification of the vector of link characteristics $z_e$ and the corresponding parameter estimates are given in SI Appendix Table \ref{tab:model}.

Figure \ref{fig:NetworkCostRates} shows the estimated link cost rates on a map of the network. The main bicycle network is clearly visible and seems to be quite dense and well connected.  Such maps may be used by urban planners to suggest areas where the bicycle network could be improved. The specification of the cost rate comprises variables that indicate the type of infrastructure for each link in the network, including information about the type of bicycle infrastructure, and the nearby land use.  We discuss the most important parameters in turn. 

The reference category is residential roads without bicycle infrastructure in low-rise urban areas. 
Compared with the reference, bicyclists associate a 11\% higher cost rate with large roads (roads with at least two lanes in one direction), whereas the difference to medium roads (roads with at most one lane in each direction) is small and statistically insignificant.

Provision of dedicated bicycle infrastructure  quite substantially reduces the subjective cost of bicycling. Cycleways (bicycle paths in own trace) reduce the subjective cost by 20\%. On residential and medium roads, bicycle lanes, whether protected or just painted, reduce the cost rate by 14\% and 22\%, respectively. The type of bicycle lane has a considerable effect on the cost rate for the large roads category: painted bicycle lanes have only a small and statistically insignificant effect, whereas protected bicycle lanes reduce the cost rate by 34\%. It makes clear intuitive sense that the impact of bicycle lanes is larger, the larger the road is, and only protected lanes affect the largest roads where car traffic is heavier.

A number of routes are branded as so-called cycle superhighways. This is a label given to high-quality, continuous bicycle routes, that cater to commuter cyclists \textcite{liu2019practitioners}. Additional routes are planned to become cycle superhighways in the future, but have not yet received the label \textcite{Visionsplan2017}. We estimate a cost rate reduction of 12\%, both for the actual and the planned cycle superhighway links. This suggests that the routes included in the cycle superhighway network were ex ante attractive and that the transformation from planned to actual cycle superhighway does not yield any additional cost reductions beyond those already accounted for at the link level.  

The model also includes parameters accounting for interactions between the type of infrastructure and the neighboring land use. 
The cost rate is much reduced for cycleways in industrial areas (48\%) or green areas (53\%). It makes intuitive sense that cycleways in green areas may be pleasant. Another potential explanation which also applies for industrial areas is the attractiveness of isolation from heavy traffic. 

In summary, provision of bicycle-friendly infrastructure has a substantial effect on route choice. We shall see below that this translates into a substantial effect on the number of bicycle trips.



\subsection{Bicycle travel demand} \label{sec:BicycleTravelDemand}

We set up a gravity model to measure the association between the number of bicycle trips in each OD pair and the characteristics of the bicycle network. The route choice model parameter estimates show that the characteristics of the network significantly affect the subjective cost of using the links of the network. Therefore, the route choice model can be used to compute the subjective cost of traveling by bicycle in any given OD pair, thereby aggregating the network information in a model-consistent manner.  

Let $\hat c ^{od} =  \sum_{e \in \mathcal{E}} l_e \left(c_e \hat x_e^{od} + F(\hat x_e^{od}) \right)$ be the subjective cost associated with the cost-minimizing flow $\hat x ^{od}$ connecting origin $o$ to destination $d$ and let $Y^{od}$ be the observed number of trips in the OD relation $od$. We assume that $Y^{od}$ follows a Poisson distribution with expectation given as a log-linear function of the cost:
    \begin{equation}
    \ln E \left[Y^{od}\right] = D\left(\hat c^{od}\right) + \delta + \eta_o + \gamma_d,     
    \label{eq:Gravity}
    \end{equation}
where $\delta$ is a constant, and $\eta_o$ and $\gamma_d$ are constants for each origin and destination, respectively, except one. The constants account for the total bicycle traffic volume out of each origin and the total volume into each destination. The demand function $D$ is expected to be downward sloping, such that higher cost implies less volume. We specify $D$ to be continuous and piecewise linear with the number of pieces chosen by eye-balling. 

\begin{figure}[b!]
    \centering
    \includegraphics[width=\columnwidth]{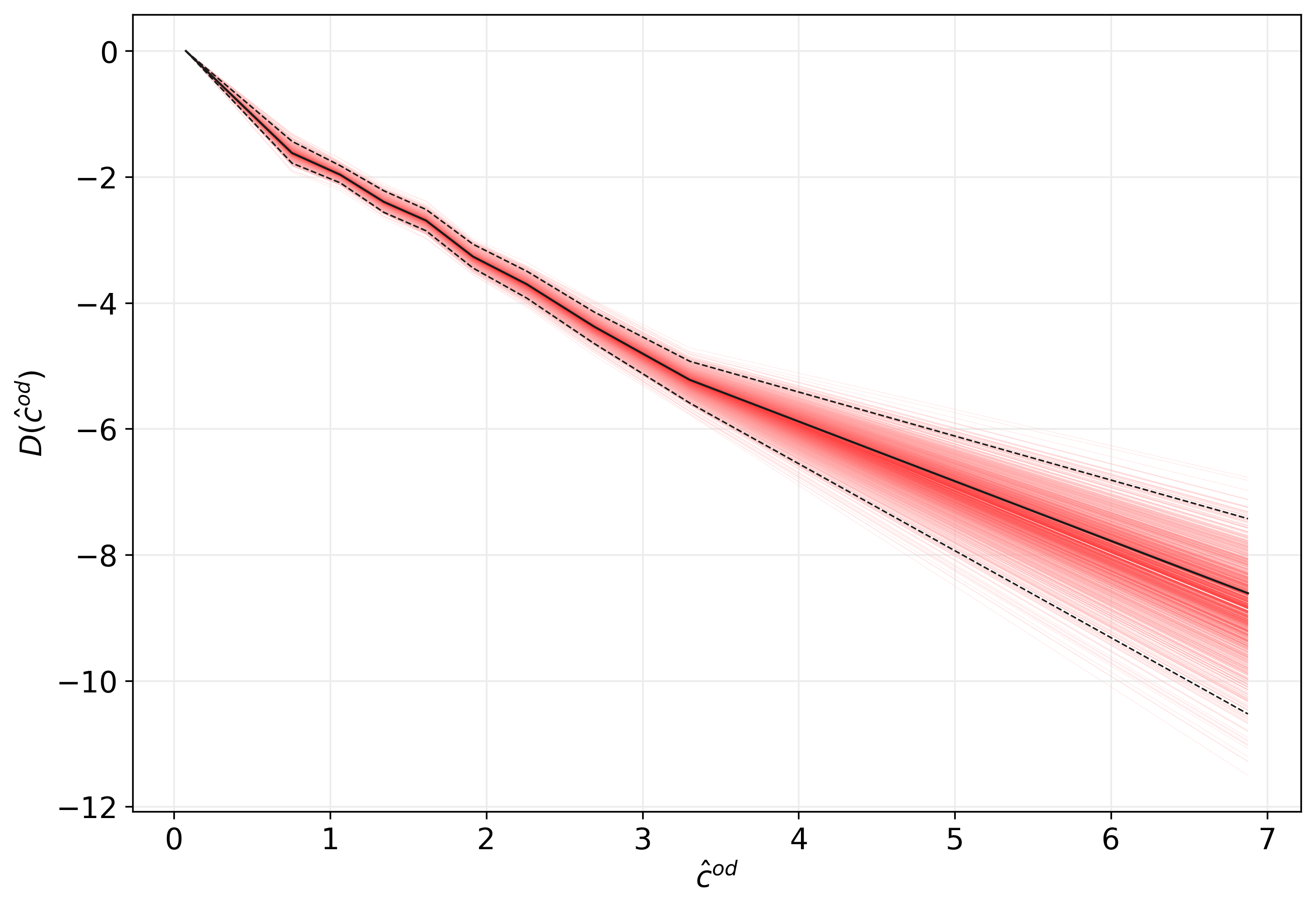}
    \caption{Piecewise linear estimate of $D$ using 10 knots. n = 38,649, $R^2=0.43$, and  $\text{\textit{RMSE} (in levels)} = 10.6$. Red lines trace pointwise confidence intervals at various quantiles computed using 1,000 bootstraps. Black dashed lines indicate the estimated 95\% pointwise confidence band. }
    \label{fig:dFunction}
\end{figure}

Figure \ref{fig:dFunction} shows the estimated demand function. The figure also shows a bootstrapped pointwise confidence band for the estimate of the demand function. The confidence band is very tight at small values of $\hat c ^{od}$ and is wider at larger values where observations are fewer. 
From the estimated demand function, we can compute the implied demand elasticity as a function of the cost. We find that the demand elasticity decreases almost linearly from 0 to about -6.5 as the cost increases from 0 to 7. A 10\% increase in the cost of a short trip thus has very little impact on the number of bicycle trips while it reduces the number of trips by up to 65\% for the longest trips.

\FloatBarrier

\subsection{Counterfactuals}
We now exploit our model to simulate the impact of general counterfactual changes to the bicycle-relevant network to illustrate how much bicycle infrastructure has contributed to encouraging bicycling in  Copenhagen. These results may be of interestfor other cities aiming to improve or expand their bicycle-relevant network.  

Table \ref{tab:scenarios} summarizes the results for the counterfactual scenarios. We compute the change in consumer surplus for bicyclists as well as the change in external cost owing to health and accidents. We have scaled the gravity model output such that the base scenario reflects the total number of kilometers traveled by bicycle in the region. A full economic evaluation of constructing bicycle infrastructure would also need to take into account construction costs, the effects on travel times by car, and the induced effects on climate, accidents, noise, and air pollution.

\begin{table*}[t!] \footnotesize
\centering
\resizebox{\textwidth}{!}{\begin{tabular}{l|rrr} 
\hline
 &  & \multicolumn{1}{l}{\textbf{Painted lanes instead of pro-}} & \multicolumn{1}{l}{\textbf{No existing \& planned cycle}} \\
\textbf{Scenario} & \multicolumn{1}{l}{\textbf{No bicycle lanes}} & \multicolumn{1}{l}{\textbf{tected lanes on large roads}} & \multicolumn{1}{l}{\textbf{superhighway classifications}} \\
\midrule
\textbf{Length of network  changed [km]} & \tabsingle{1,427.8} & \tabsingle{407.0} & \tabsingle{*333.7}\\
\textbf{Avg. bicycle travel cost [rel. increase,\%]} & \tabsingle{34.3}
 & \tabsingle{7.1} & \tabsingle{12.8} \\
\textbf{\# trips [relative decrease,\%] } & 
\tabtrip{37.0}{35.1}{38.6} & \tabtrip{9.9}{7.2}{12.1} & \tabtrip{16.8}{14.3}{18.8} \\
\textbf{Total distance traveled by bicycle [relative decrease,\%]} & 
\tabtrip{46.8}{44.9}{48.7} & \tabtrip{13.7}{10.9}{16.2} & \tabtrip{22.8}{20.2}{25.2} \\
\midrule
\textbf{Loss of consumer surplus [M € / year]} &
\tabtrip{174.1}{171.8}{176.2} & \tabtrip{44.8}{44.1}{45.5} & \tabtrip{75.6}{74.6}{76.6}\\
\textbf{Loss owing to health and accidents [M € / year]} &
\tabtrip{435.3}{418.1}{452.9} & \tabtrip{127.8}{101.1}{150.9} & \tabtrip{212.5}{188.3}{234.4} \\
\textbf{Societal loss [M € / year]} & 
\tabtrip{609.4}{594.3}{624.9} & \tabtrip{172.6}{146.7}{195.0} & \tabtrip{288.1}{265.0}{309.0} \\
\textbf{Loss relative to length of network changed [k € / year / km]} & 
\tabtrip{426.8}{416.2}{437.7} & \tabtrip{424.0}{360.5}{479.0} & \tabtrip{*863.6}{794.1}{926.0} \\
\bottomrule
\end{tabular}}
 \caption{Key figures of three counterfactuals. Numbers in square brackets indicate 95\% bootstrap confidence intervals based on 1,000 full sample repetitions. *Measured as route distance rather than  network distance.}
 \label{tab:scenarios}
\end{table*}

The first counterfactual simulates a situation where all 1,427.8 km of bicycle lanes and all cycle superhighway classifications have been removed. On average, this increases the subjective cost of bicycling by 34.3\% per km, which induces a decrease  of 37.0\%  in the number of bicycle trips and 46.8\% in the total distance traveled by bicycle. The bootstrapped confidence intervals indicate that these numbers are quite precisely determined. Relative to the situation without bicycle lanes, the simulation thus suggests that the provision of bicycle lanes has induced an increase in bicycle use by 59\% (trips) and 88\% (total distance traveled). 

To measure the loss to bicyclists in the counterfactual scenario compared with the base scenario, we have computed the change in the consumer surplus \textcite{Mas-Colell1995}.  To convert this number from subjective cost units to monetary values, we first apply the sample average speed to convert the subjective cost to time units, and then apply the official Danish value of travel time \parencite{Enhedspriser2022} to convert from time units to monetary units. Our results suggest a decrease in consumer surplus of €174.1M per year. 

Bicycling is associated with both health benefits and accident risk \parencite[e.g., ][]{DeHartog2010,Oja2011}. The official Danish guidelines for cost-benefit analysis suggest a net external benefit owing to health and accidents of 0.91 EUR per bicycle km \parencite{Enhedspriser2022}. Applying this figure, we estimate the welfare loss induced by removing the bicycle lane network through health and accidents to be €435.3M per year. In total we find a loss of €609.4M per year or €0.427M annually per km of bicycle lane if all bicycle lanes were removed. 

In the second counterfactual, we convert the 407.0 km of protected lanes on large roads to painted lanes, while maintaining the cycle superhighway classifications. This increases the subjective cost of bicycling by 7.1\% on average, which in turn induces 9.9\% less bicycle trips and 13.7\% less kilometers travelled by bicycle. We find that downgrading protected lanes leads to an annual loss of €44.8M of consumer surplus and an annual loss due to health and accidents of €127.8M. In total, we compute an annual loss of €172.6M€ or €0.424M per km of protected bicycle lane.  

The third counterfactual removes the existing and planned cycle superhighway classifications. We interpret this as representing the effect of no longer having long, connected bicycle routes. The change involves 333.7 km of cycle superhighway routes and leads to an average increase of 12.8\% in the subjective cost, which induces a 16.8\% decrease in the number of trips and 22.8\% decrease in the number of kilometers traveled by bicycle. Removing the route-level features that constitute the cycle superhighways is associated with a annual total loss of €288.1M or €0.864M  per lane km. 

In all three counterfactuals, we find that the total distance traveled by bicycle responds relatively more than the number of trips. This means that the number of long bicycle trips responds more than the number of short trips, in line with our observation that the demand elasticity increases with the trip cost.

\FloatBarrier

\section{Discussion}

We find substantial impact of the provision of bicycle-relevant infrastructure on the subjective cost and the volume of cycling. We work at a very fine level of resolution, which allows us to distinguish between a large number of infrastructure types. This is first-order important, as we find a difference in the subjective cost of cycling of more than a factor eight between the best and the worst infrastructure types. Thus, the type and the location of infrastructure are very important. 

The counterfactual simulations performed in this study illustrate the effect of broad changes to the bicycle network. These results may be of interest when considering the consequences of expanding the bicycle network in cities with less bicycle infrastructure than Copenhagen. We find that the existing bicycle network in Copenhagen has led to a substantial increase of about 90\% in distance traveled by bicycles. These changes can be interpreted as representing short-term effects, as they hold constant the fixed effects associated with origins and destinations. In the longer term, location patterns can be expected to adapt to improvements in the bicycle network, making the long-term effect of a network improvement larger than the short-term effect. Previous research supports the broad conclusion that bicycle infrastructure induces more bicycle traffic \textcite[e.g., ][]{Pucher2009,buehler2016bikeway,Aldred2019a,Kraus2021}.  

From the counterfactual scenarios, we have calculated the net benefit of bicycle lane  provision associated with the change in subjective cost, health, and accidents to be €420k--440k per lane km per year. According to \parencite{ECF2021}, construction costs are in the range €0.5M--1.5M per lane km.\footnote{Lower if the numbers in \textcite{ECF2021} pertain to route km.} The estimated benefit associated with cycle superhighway status is greater, €860k per km per year, although it relates only to route-level features, holding link-level features constant.  As construction costs are incurred once but benefits accrue year by year, these results indicate that the provision of well-located and high-quality bicycle infrastructure can easily generate a positive net present value in a standard cost-benefit analysis. 

Copenhagen already has extensive bicycle infrastructure, so the effect of additional infrastructure may be smaller. On the other hand, we find a large net benefit of cycle superhighways, which may arise from having long and connected bicycle routes. This means that  limited investments can potentially lead to  large net benefits by improving overall connectivity of the bicycle network. Maps such as Figure \ref{fig:NetworkCostRates} can be used to identify candidate locations for such investments. 

We have combined a very large database of observed bicycle trajectories and a very fine-grained representation of the bicycle-relevant network with a modeling approach that allows us to take the entire network into account. Our model can be applied to predict the effect of providing specific infrastructure in specific places. Similar analyses can be undertaken for other cities. In such analyses, however, the main obstacle is obtaining sufficient data on observed route choices similar to the Hövding dataset used in this study. 

This is the first study of its kind, so there is much scope for future research. On the bicycle front, it is of interest to estimate similar models using datasets from other cities to consolidate and extend the conclusions regarding the impact of bicycle infrastructure on bicycle demand. Another research avenue is investigating datasets sampled by different means or from different kinds of users to check the robustness of our conclusions. 

On the methodological front, a general research agenda can be formulated for the perturbed utility route choice model, with a view to applications to bicycle traffic or other traffic through complex networks. The most important point here, we think, is to develop approaches that allow the estimation of the models at the level of individual trajectories. This would make it possible to avoid the data loss associated with the aggregation of data to the OD level and allows the inclusion of individual-level information. Another related avenue is to develop solution methods for the cost minimization problem in \eqref{eq:U} that make it feasible to work with meta-networks where link-pairs take the place of links. This would allow turn movements to be represented and hence allow to take into account, for example the cost of left turns and crossing roads with car traffic.


\section{Materials and methods}
 \subsection{Bicycle data}

\begin{figure}[h]
    \centering
    \scalebox{1}[-1]{\includegraphics[width=0.5\textwidth]{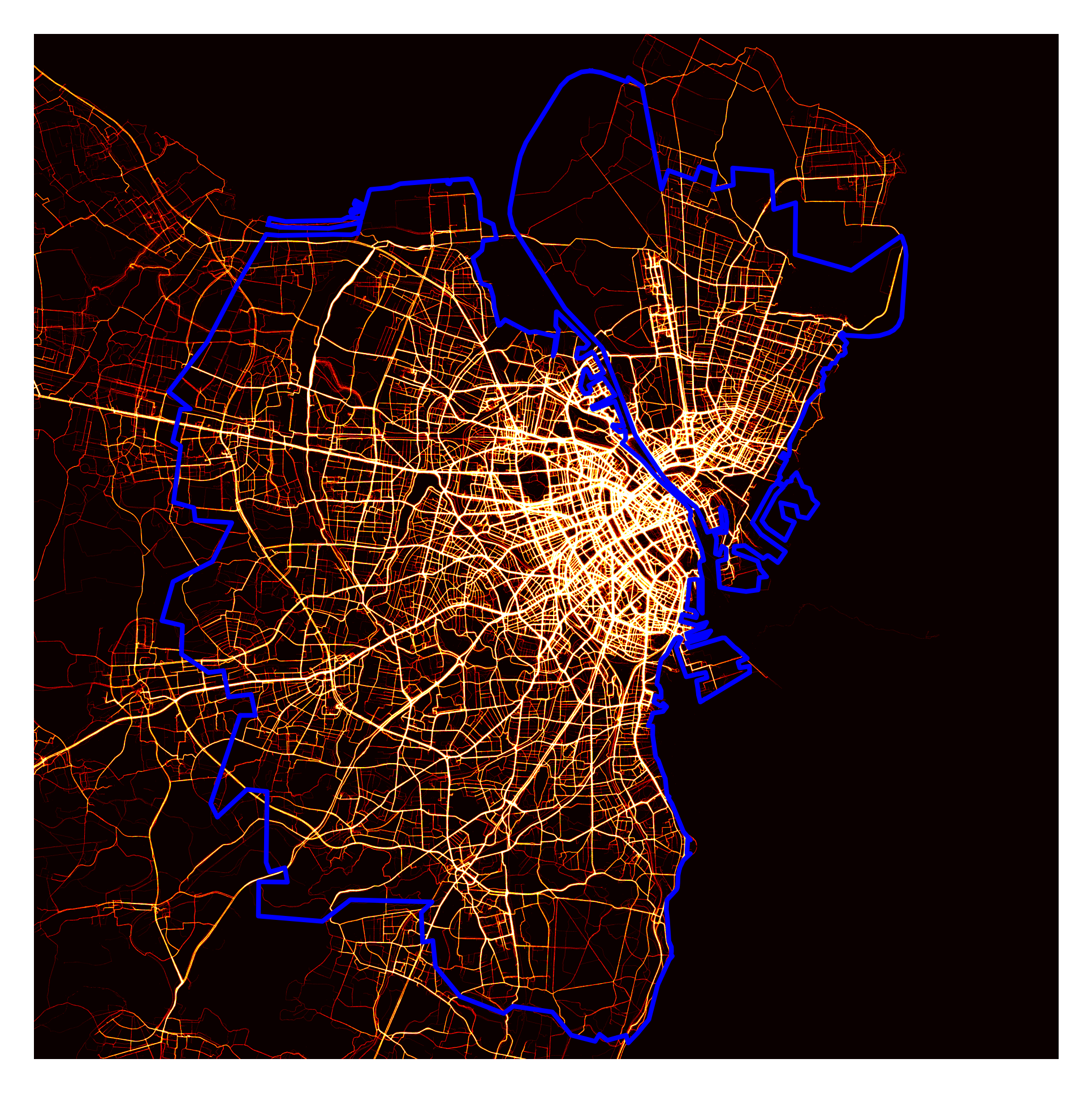}}
    \caption{GPS traces of bicycle trips in Greater Copenhagen}
    \label{fig:GPStraces}
\end{figure}

Figure \ref{fig:GPStraces} shows the raw data from our sample of GPS traces of bicycle trips, collected in Greater Copenhagen \parencite{Hovdingdata}. Figure \ref{fig:bicyclenetwork} presents the corresponding views of the road and dedicated bicycle network, using Open Street Map data \parencite{OpenStreetMapdata}. Copenhagen has an extensive bicycle network with many cycleways, especially outside central Copenhagen. In central Copenhagen there is a dense network of protected bicycle tracks and many non-protected bicycle lanes.\footnote{See SI Appendix  \ref{Sec:SI data and data processing}\ref{sec:SI Network data} for definitions of the various infrastructure types.}

SI Appendix \ref{Sec:SI data and data processing}\ref{sec:Data:MapMatching}-\ref{sec:Data:OD} describes how we processed our data. In brief, the pre-processed dataset of  218,489 GPS trajectories collected from 8,588 individuals were map-matched to the Copenhagen bicycle network using software presented in \textcite{Haunert2012}. Our estimator for the route choice model requires trips to be aggregated such that each included OD pair has at least two distinct observed trajectories. Therefore, we applied an algorithm that trims individual trajectories at both ends such that the trimmed trajectories have a small number of origins and destinations in common. Our analyses are based on data with 200 origins and 200 destinations, which represents a compromise between including most of the observed trajectories and avoiding many OD pairs with only a small number of observations. Robustness check with 100 and 400 origins and destinations did not indicate problems; see SI Appendix Section \ref{sec:S route choice model}\ref{sec:ChangingNumberOfODs}. We use the data for all OD pairs that are more than 1 km apart and have at least two distinct individual route choice observations per OD. Our estimation data comprise 152,323 trimmed trips from 7,672 individuals.  

\subsection{Route choice model}

A directed network comprising nodes and links $(\mathcal{V},\mathcal{E})  $ is described by the incidence matrix $A$ with elements $a_{ve}=1$ if $v$ is the origin node of link $e$, $-1$ if it is the destination node, and $0$ otherwise. A set of OD pairs $\mathcal{B}$ is represented in terms of OD demand vectors $b \in \mathcal{B}
\subset  \mathbb{R}^{|\mathcal{V}|}$,
where $b_v = 1$ indicates the origin node of trip $b$,  $b_v = -1$ indicates the destination node and $b_v = 0$ otherwise. 
The flow conservation constraint $Ax=b$ ensures that a non-negative flow vector $x \in \mathbb{R}_+^{|\mathcal{E}|}$ is physically consistent with demand $b$ through the network.


The perturbed utility route choice model holds that the flow vector $\hat x ^b$ for bicyclists with demand $b$ minimizes the cost function in \eqref{eq:U} under the flow constraint $A\hat{x}^b=b$. Fosgerau et al. \textcite{Fosgerau2021a} show that this model generates very reasonable substitution patterns. Moreover, the model directly applies to the complete network, without a need to specify a choice set of route alternatives.  

Given (noisy) observations of flow vectors, Fosgerau et al. \textcite{Fosgerau2021a} transform the active first-order conditions for the cost minimization problem to a linear regression equation that directly leads to an estimate of $\beta$. There is an active first-order condition for each link with positive observed flow. The transformation eliminates Lagrange multipliers corresponding to the flow conservation constraints at each node of the network. The data for the regression comprises many observations for each OD pair, which enables standard errors to be clustered by OD pair.

\begin{figure}[b!]
    \centering
    \includegraphics[width=0.49 \textwidth]{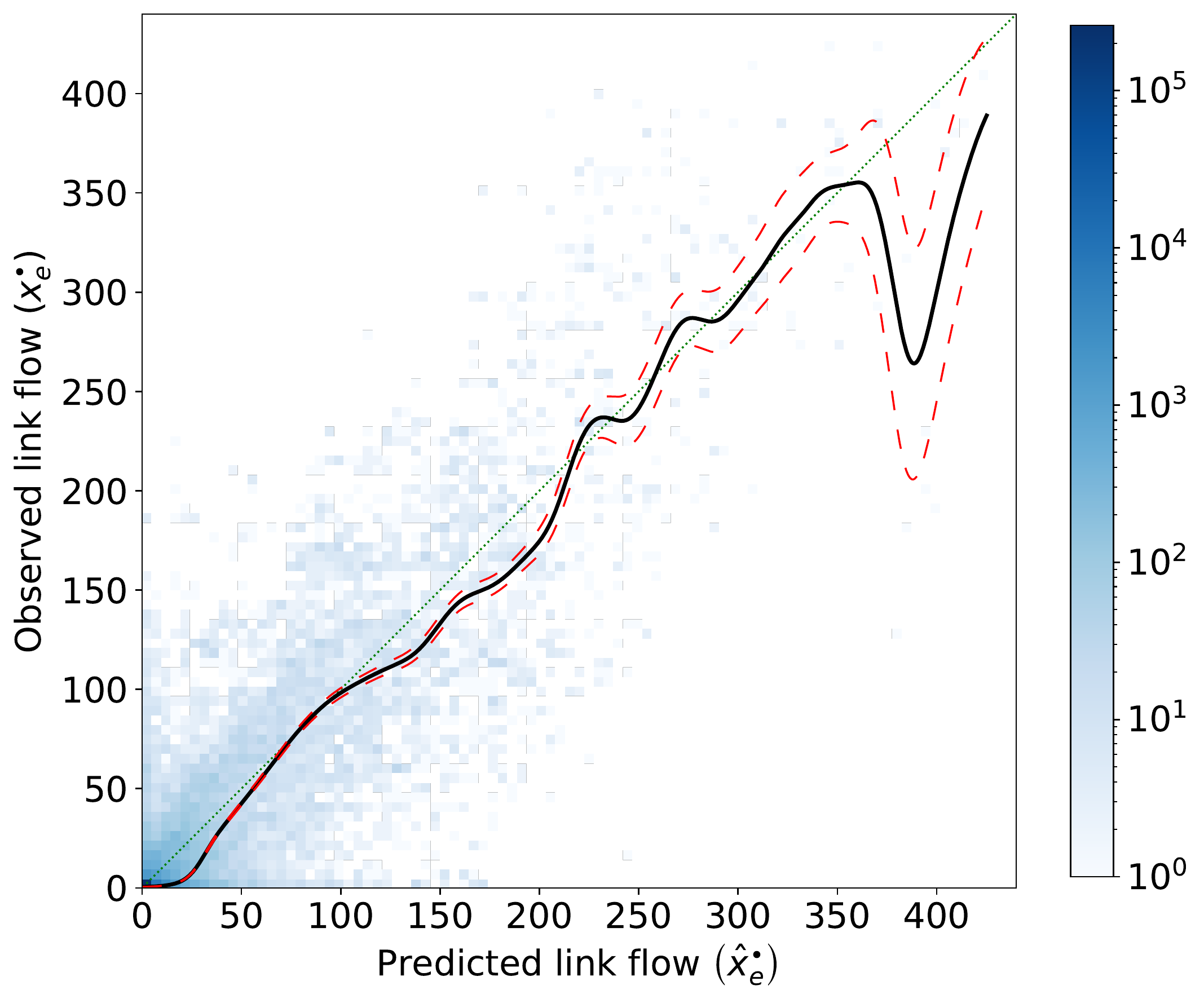}
    \caption{Heatmap of total observed link flow $\left(x_e^{\bm\cdot} = \sum_{o\in \mathcal{O}} \sum_{d \in \mathcal{D}} x_e^{od}\right)$ against the total predicted link flow $\left(\hat x_e^{\bm\cdot} = \sum_{o\in \mathcal{O}} \sum_{d \in \mathcal{D}} \hat x_e^{od}\right)$ for each link $e \in \mathcal{E}$. The color of each grid cell represents the number of links belonging to each cell. The thin green dotted line is the $45^\circ$ line.
    The black line is a Nadaraya-Watson non-parametric regression \parencite{Nadaraya1964,Watson1964} with Gaussian kernel and bandwidth $10$ chosen by eyeballing. The corresponding 95\% pointwise confidence band is indicated by dashed red lines. }
    \label{fig:ValidationPartCStyle}
\end{figure}

Figure \ref{fig:ValidationPartCStyle} plots the total observed link flow ($x_e^{\bm\cdot} = \sum_{o\in \mathcal{O}} \sum_{d \in \mathcal{D}} x_e^{od}$) against the total predicted link flow ($\hat x_e^{\bm\cdot} = \sum_{o\in \mathcal{O}} \sum_{d \in \mathcal{D}} \hat x_e^{od}$) for each link $e \in \mathcal{E}$ across all origins and destinations. A perfect prediction would exactly follow the $45^\circ$ line. We find that a non-parametric regression line quite closely tracks the $45^\circ$ line. This is satisfactory, especially considering that the route choice model uses only 28 parameters. The correlation (defined in the SI Appendix \eqref{eq:ValidationCorrelation}) between $x^{\bm\cdot}$ and $\hat x^{\bm\cdot}$, is $ \rho(x^{\bm\cdot}, \hat x ^{\bm\cdot}) = 0.8894$.

SI Appendix \ref{sec:S route choice model}\ref{sec:Validation} comprises a range of validation tests of the route choice model.

\section*{Acknowledgements}{This research is funded by the European Research Council (ERC) under the European Union's Horizon 2020 research and innovation programme (grant agreement No. 740369). We thank Michel Bierlaire, Mike Smith, seminar participants at the Tinbergen Institute in Amsterdam, conference audiences at the 7\textsuperscript{th} International Choice Modelling Conference in Reykjavik, and the 10\textsuperscript{th} Symposium of the European Association for Research in Transportation (hEART) in Leuven for comments.}

\printbibliography[title=References]

\newpage

\appendix

\section{Literature review}
    
\subsection{Perturbed utility} \label{sec:Lit:PURC}
In a general perturbed utility model, a consumer chooses a consumption vector $x$ from some budget set $B$ that solves a utility maximization problem of the form $\hat x= \argmax_{x \in B}(a^\intercal x - F(x))$ --- that is where the utility function is a linear function ``perturbed'' by subtracting a convex function \parencite{McFadden2012, Fudenberg2015,Allen2019}. Perturbed utility models are firmly rooted in modern microeconomic theory and can be interpreted as representing a population of optimizing agents whose individual behavior is described by one of a wide range of models \textcite{Allen2019c}. The additive random utility discrete choice model \textcite{McFadden1981} belongs to the set of perturbed utility models in which the budget set $B$ is the set of probability vectors. 

\textcite{Fosgerau2021a} introduced the perturbed utility route choice (PURC) model. The model represents traveler behavior as a utility-maximizing flow across an entire network. The perturbed utility budget now requires that flow is conserved through all network nodes from origin to destination. This is a partial specification of behavior because it does not specify exactly which route a given traveler will take but only specifies the probabilities that are implied by the optimally chosen flow. A special perturbation function is specified that induces a tendency to distribute flow on more routes while allowing the optimal flow to be zero on most links in the network. \textcite{Fosgerau2021a} estimate and validate the model using 1,337,096 car trips in a large road network.

\subsection{Bicycle route choice models}
The bicycle route choice literature has almost exclusively relied on path-based models. These are random utility discrete choice models that view the route choice as a discrete choice among a  predefined set of path alternatives \parencite{menghini2010route,hood2011gps,Broach2012a,khatri2016modeling,ton2017people,casello2014modeling,chen2018gps,Ghanayim2018,Prato2018,sobhani2019metropolis,fitch2020road,shah2021different,cho2022estimation}. However, predefining the choice set is problematic, as it leads to bias in the parameter estimates \textcite{frejinger2009sampling} and risks excluding the actually chosen alternatives. This is a problem, in particular, for bicycle route choice, as bicycle networks tend to be very fine-grained \parencite{ton2017people}. Researchers have been forced to discard almost half of the observed trips owing to inadequate similarities of observed trips with any of the predefined alternatives \parencite{koch2021limitations}.

The recursive logit model \parencite{Fosgerau2013e} and the nested recursive logit model \parencite{Mai2015} view the choice of a path as a Markov chain of link choices, where the traveler at each step anticipates the expected utility associated with reaching the next node. The recursive logit models incorporate the entire network and thereby avoid pre-defining choice sets. However, a downside of this feature is that the models distribute positive flow on all network links, whereas in reality most of the network will be unused for any given origin-destination (OD) pair. 
\textcite{Zimmermann2017} have applied recursive logit models to bicycle route choice, but the computation time was 15 days for the nested recursive logit model and 43 hours for the recursive logit model. \textcite{koch2021limitations} were not able to obtain identification with recursive logit models. An inherent limitation of these models is that computation times are strongly affected by the size of the network. The network in \textcite{Zimmermann2017} has $42,000$ links, so the estimation of a recursive logit model for the network with  $420,000$ links used in the present study would most likely not be feasible.  

In this study, we instead model bicycle route choices using the PURC model, which has a range of attractive features. The model does not require choice sets to be pre-specified; it simply incorporates the entire network. In contrast to the recursive and nested recursive logit models, the PURC model leaves most of the network unused for any OD. The model allows very fast estimation of link cost parameters using linear regression, even when considering a large number of link-level variables. The model also allows estimation on much larger datasets (we estimate our model on more than 150,000 trips within a minute). Our estimating method using the PURC model requires the pre-processing of data, which leads to some loss of data; nevertheless, we are able to retain 70\% of the observed route choices (see Section \ref{Sec:SI data and data processing}\ref{sec:Data:OD}), which is more than the choices retained with path-based approaches.

\subsection{Bicycle infrastructure and the demand for bicycling}


\textcite{Pucher2009,buehler2016bikeway,Aldred2019a} review the empirical evidence on the effect of bicycle infrastructure on bicycle demand. The evidence supports that the infrastructure mostly does affect cycling demand and that the effect is heterogeneous across infrastructure types \textcite{Pucher2009}. \textcite{Aldred2019a}  points out that previous studies are mostly case-based, and \textcite{buehler2016bikeway} calls for studies that use individual-level data to assess the influence of the entire network on bicycle demand. 


Going into the individual studies, one group of studies reviewed by \textcite{VanGoeverden2015} and \textcite{Molenberg2019} conducts case-based before-and-after analyses of the demand effects of specific new bicycle infrastructure projects, using traffic count data or travel surveys \textcite[e.g., ][]{Merom2003,Evenson2005,Krizek2009,Goodman2014,skov2017effects, hong2020evaluation}. Such studies are directly aimed at identifying the causal effects of new infrastructure. They find that demand increases  up to 140\% on the specific new pieces of infrastructure. Owing to the research design, however, these studies cannot distinguish between traffic actually induced by the new infrastructure and existing traffic that is attracted from elsewhere \parencite{Molenberg2019}. In the present study, we clearly distinguish between these two mechanisms. These studies are also less suited for identifying the effects of the specific attributes of the new infrastructure, which the present study finds to be very important.

A second group of studies relates bicycle demand to the attributes characterizing entire networks at the macro-level \textcite{Nelson1997,Dill2003,Pucher2005,Parkin2008,schoner2014missing, osama2017models,Nielsen2018,Ryan2020,Vandenbulcke2011,fagnant2016direct}. 
These studies are cross-sectional, either at the city or country level \parencite{Nelson1997,Dill2003,Pucher2005,schoner2014missing,Vandenbulcke2011} or at the level of areas within a  region \parencite{Parkin2008,fagnant2016direct,osama2017models,Nielsen2018,Ryan2020}.
The general findings from these studies are that the overall bicycle demand is positively correlated with the length of the bicycle network \parencite{Nelson1997,Dill2003,Pucher2005,Parkin2008,Vandenbulcke2011,osama2017models,Nielsen2018} and is negatively correlated with the share of large roads \parencite{osama2017models,Vandenbulcke2011}. 
Both findings are supported by our study, in which we further find that the preference against large roads more than disappears when the roads are equipped with protected bicycle lanes.

\textcite{Kraus2021} consider the effects of provisional pop-up bicycle lanes on travel demand during the COVID-19 pandemic in a cross-section of European cities. They find that the pop-up bicycle lanes induced an average increase in cycling of 41.6\%, some of which may be new bicycle traffic while some may just be diverted. The effect of pop-up lanes was lower in cities with a larger pre-existing bicycle network per capita. For comparison, in our counterfactual simulation wherein all bicycle lanes are removed, we find that the number of bicycle trips decreases by 29.9\%. The research design in \textcite{Kraus2021} arguably allows the identification of causal effects but cannot account for network-wide effects or explicitly for the characteristics of the existing infrastructure.

A third group of studies relates the bicycle modal share at the level of OD pairs to corresponding distances based on shortest paths \parencite{iacono2008access,lovelace2017propensity}. Their results broadly agree  with ours, but in contrast to the present study, they cannot take the quality of the bicycle-relevant network into account. 

Finally, a fourth group of studies uses traditional transport models \parencite{Liu2020,hallberg2021modelling,Rich2021} or agent-based transport simulation  \parencite{Agarwal2019} to model bicycle flows in a  network. This allows them to analyze detailed counterfactual scenarios without having access to observed route choice data. \textcite{hallberg2021modelling,Agarwal2019,Rich2021} found an increase of 18\%-35\% (Patna, India) and 4\%-9\% (Copenhagen, Denmark) in bicycle use resulting from  building/expanding the network of cycle superhighways. In our study, we find a similar effect in a counterfactual where the number of bicycle trips drops by 16.8\% when the existing and planned cycle superhighways are removed.  \textcite{Liu2020} finds a generalized cost elasticity of demand of $-0.7$, whereas the present study finds a higher elasticity in the range $(-1.39,-1.08)$. Equipping all links with bicycle lanes leads the demand to increase by 20\% in \textcite{Liu2020}, whereas we find demand to decrease by 37\% when we remove all bicycle lanes. 

In this study, we estimate a combined bicycle route choice model and demand model. Our model is firmly rooted in data; it incorporates a very large dataset comprising more than 150,000 observed bicycle trips across a large network and includes an extensive set of explanatory variables. From the observed route choices, we determine a generalized cost measure that is used to predict bicycle demand across the network. This improves on previous studies on a number of points. In particular, we can incorporate all bicycle infrastructure in the network and not just a few specific cases of new infrastructure. We can also assess the effects of detailed and network-wide counterfactual changes to the bicycle network, distinguishing between new and diverted bicycle trips. Our generalized cost measure integrates all our observed infrastructure attributes across the network to the extent that they affect route choice. The model can thus take into account both the quality of bicycle infrastructure and its location.

\section{Data and data processing}\label{Sec:SI data and data processing}

\subsection{Network data}
\label{sec:SI Network data}
The network representation is based on Open Street Map \parencite[OSM,][]{OpenStreetMapdata} and includes the bicycle-relevant infrastructure --- that is all network links where riding or carrying a bicycle is possible, including the elements listed in Table \ref{tab:attr}. In the representation, bicycling in both directions is allowed on all network links, while keeping track of the direction. The resulting network representation  of the Copenhagen Metropolitan Area contains a total of 420,973 directed links and 324,492 nodes. 

We define infrastructure types by combining three infrastructure attributes: road type (based on OSM tags), road size (based on the number of car lanes), and type of bicycle infrastructure (whether it is present and if so, whether it is a protected or a painted bicycle lane). This creates 16 distinct infrastructure types, as shown in Table \ref{tab:attr}.

The OSM network attributes have been enriched with information on land use and elevation. The land use information is obtained from an external shapefile layer \parencite{fot-kort10_fot-kort10_2018}, and includes the following categories: green areas (including green restricted areas, parks, and forests), areas near water, industrial areas, open landscape areas, low-rise urban areas, and high-rise urban areas (merged with the city center). For  each directed link, the land use on the immediate right-hand side of the link is determined, tracking the length of each land use category. The elevation gradient is computed with 10 m splits of the network. Using overlay analysis, elevation information per 10 m is attributed to each link. Based on this, the slope and difference in elevation are obtained per 10 m on each link, and the total vertical meters gained when the slope is greater than 3.5 \% per direction per link are determined. 

\subsection{Trajectory data} \label{sec:Data:TrajectoryData}
The data were collected in Greater Copenhagen (the study area framed in Figure \ref{fig:GPStrim}) between September 16, 2019 and May 31, 2021 from 9,564 individuals using a Hövding head protection airbag helmet designed for cyclists \parencite{Hovdingdata}. Positional data were passively collected among users who had given consent to share their data and were transmitted to a database server through the users' smartphones connected to the airbag helmet via Bluetooth. The dataset of observed trajectories contains 347,430 trips starting and/or ending in Greater Copenhagen and covering a total of 939,711.8 km traveled by bicycle. Each trip connects an OD pair, represented by the starting point and the endpoint of the trip, respectively. 

\subsection{Map matching}
\label{sec:Data:MapMatching}
Each of the observed trajectories has been map-matched to the bicycle network using the hidden Markov algorithm proposed in \textcite{Haunert2012}. The algorithm allows for off-road parts in the matched route, which are  often necessary for bicycle trips, as bicyclists do not always stick to formal roads and paths. However, our network has a high resolution, and we found that only 35 trips were matched with off-road segments. We discarded these trips from the subsequent analysis.

\subsection{OD data and trip trimming} \label{sec:Data:OD}

All trips shorter than 1 km were discarded. Furthermore,  we discarded circuitous routes that were more than $\frac{\pi}{2}$ times longer than the crow-fly distance. 
Trips with loops, where a part of the route was repeated or where the same network node was visited twice, were also discarded.
Finally, we discarded trips where the map matching algorithm failed to match the entirety of the trip.
The resulting dataset after the filtration steps comprised 218,489 trips from 8,588 individuals covering a total of 762,791.8 km.

Our estimator for the PURC model requires multiple observations in each OD pair. Because common ODs are very rare in a large network, we follow \textcite{Fosgerau2021a} and trim the observed trips such that the trimmed trips share common ODs.
Our algorithm selects first a set of origins and then a set of destinations. A trip is included in the estimation data if it passes first a selected origin and then a selected destination, and only the part of the trip between the selected origin and destination is included. 

More specifically, we include origins one by one, choosing in each step the origin that maximizes the total length of trips that it allows to include, while trimming the additional included trips to begin from that origin. After a list of origins has been compiled, we find in a similar way a list of destinations. The final output is a long list of origins and destinations. Our main results are obtained using 200 origins and 200 destinations.

\begin{figure}[ht]
    \centering
\begin{subfigure}{0.32 \textwidth}
    \scalebox{1}[-1]{\includegraphics[width=\textwidth]{FiguresPaper/Trimmed25200_Region_None_allTrips_largerCopenhagen.png}}
    \caption{After initial data filtration} \label{fig:GPStrim:a}
\end{subfigure}
    \hfill 
\begin{subfigure}{0.32 \textwidth}
    \scalebox{1}[-1]{\includegraphics[width=\textwidth]{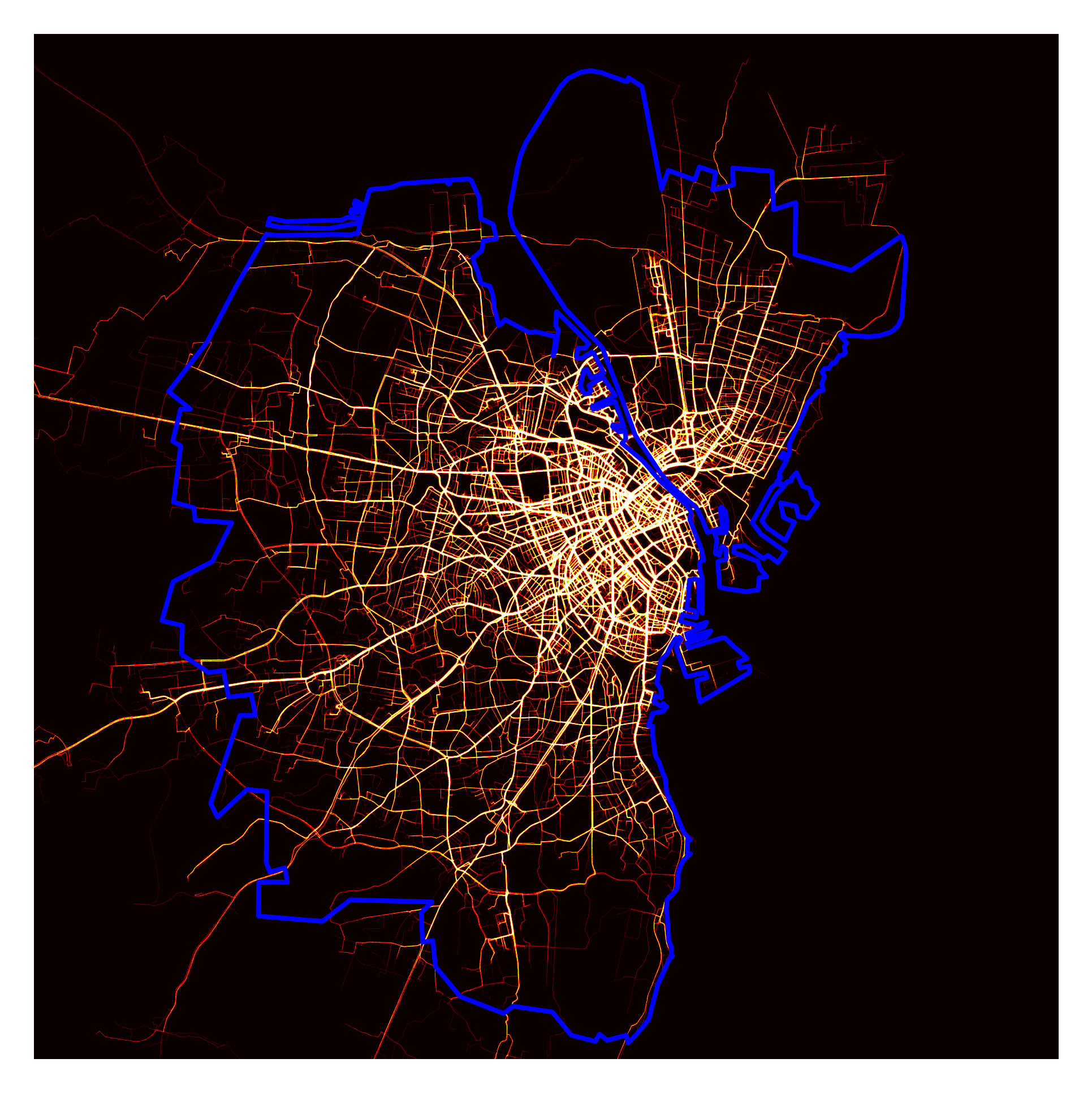}}
    \caption{Trips connecting candidate $O$s and $D$s} \label{fig:GPStrim:b}
\end{subfigure}
    \hfill 
\begin{subfigure}{0.32 \textwidth}
     \scalebox{1}[-1]{\includegraphics[width=\textwidth]{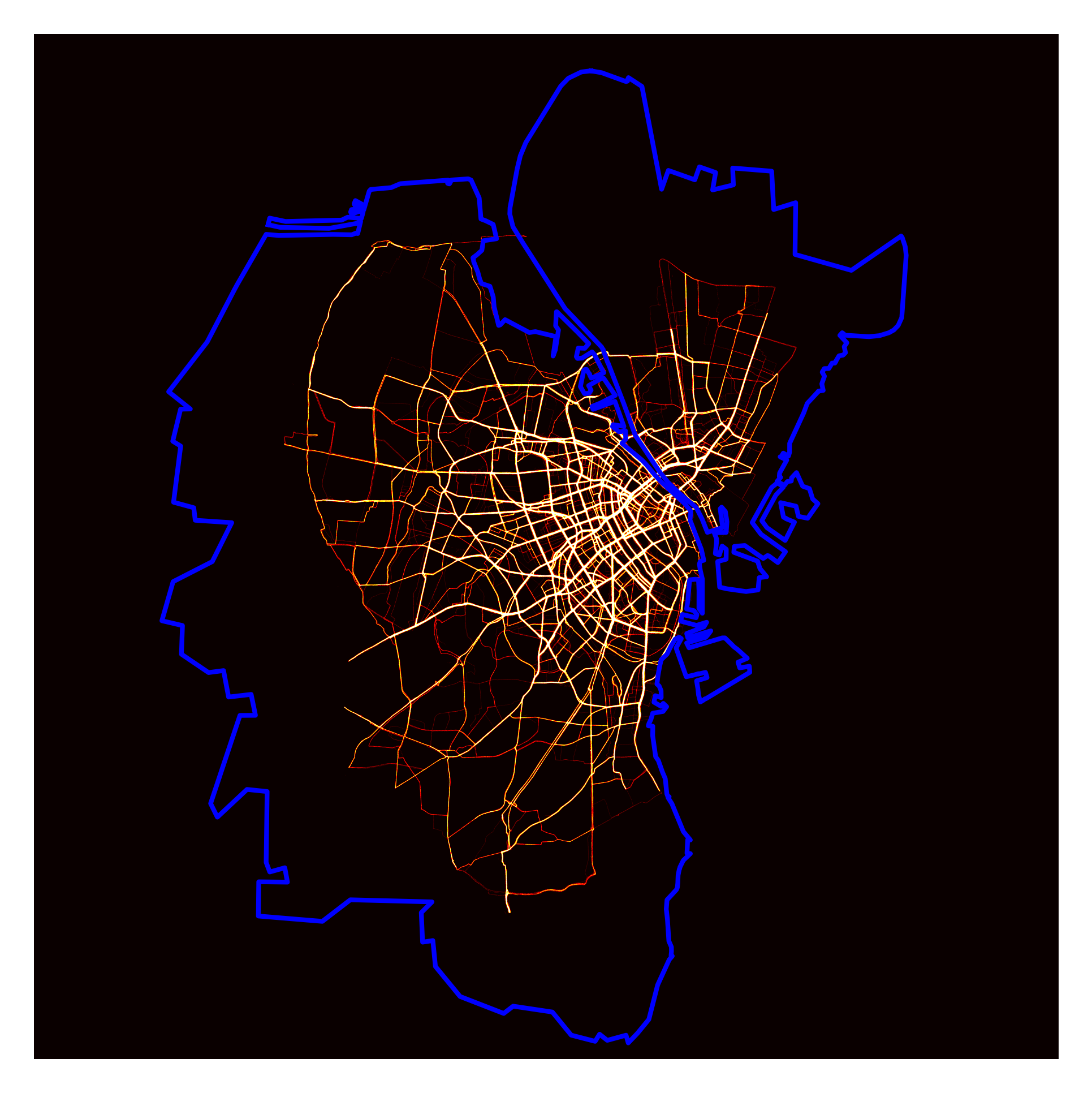}}
     \caption{Trimmed trips used for estimation} \label{fig:GPStrim:c}
\end{subfigure}
    \caption{Heat map of anonymized GPS traces after different filtration phases. 
    Study area shown in blue. (a): All GPS points after initial filtration ; (b):  GPS points of the trips used for model estimation (untrimmed filtered trips); (c): GPS points of the trips used for model estimation (trimmed filtered trips). For visualization, trips have been anonymized by removing a random number of points at their start and end.
    }
    \label{fig:GPStrim}
\end{figure}

To ensure that the generated origins and destinations are all found within Greater Copenhagen (see Figure \ref{fig:GPStrim}), in this step, we only consider trips that both start and end within Greater Copenhagen. This filtration makes the dataset used for this task slightly smaller than the final estimation dataset, where we require only that either the origin or the destination is within Greater Copenhagen (see Section \ref{Sec:SI data and data processing}\ref{sec:Data:TrajectoryData}). The dataset used for finding origins and destinations comprises 208,410 trips (703,837.5 km) across 8,456 individuals.

After compiling the list of origins and destinations, we identify, for each trip, the first origin and the last destination that are on the list. Only the trips that include first an origin and then a destination from the list are included. Included trips are trimmed to begin and end at these points.  

This process increases the likelihood of the included trips having  origins and destinations in common with other included trips. Data are lost if the number of origins and destinations is small, which speaks for including many origins and destinations. However, the estimator used combines trips that are matched using the same OD pair into an observed average flow vector for that OD pair. Increasing the number of origins and destinations means that the observed average flow will be based on fewer matched trips per active OD pair, implying more noise. It also means that there will be more unmatched trips, trips that are alone in using an OD pair, and these trips cannot be used for estimation. Therefore, we carefully choose the number of origins and destinations to balance these concerns.

At this stage, the data include trips that are not matched to another trip with the same OD. Therefore, we extend the algorithm to reduce the number of such trips. The algorithm extension first identifies the longest unmatched trip. This trip is then gradually trimmed by trying combinations of later origins (from the list of candidates) and prior destinations (from the list of candidates) until it is found to travel between an origin and a destination that matches another trimmed trip, matched or unmatched. Trips that fail to find a match are discarded. The algorithm continues with the longest remaining unmatched trip until all unmatched trips have been either matched or discarded. 

The number of origins and destinations on which to base the trip trimming was selected so as to maximize the number of OD pairs that have at least ten observed trips  (after recovering unmatched trips). This was obtained when using 200 origins and 200 destinations. 
As a check, we also report estimation results from the route choice model with 100 and 400 origins and destinations (Table \ref{tab:model100200400}). The parameter estimates are not very sensitive to this change, as we shall see in Section \ref{sec:S route choice model}\ref{sec:ChangingNumberOfODs}.

Table \ref{tab:DataFiltration} summarizes the size of the  datasets after the main steps of data processing.

In conclusion, we retained 70\% of the observed trips for the estimation. This is much higher than seen in traditional path-based route choice studies \parencite{Ton2018,Koch2021a}. 
Figure \ref{fig:GPStrim} shows the heat maps of the trajectory data after the initial data filtration (a), the trips connecting candidate origins and destinations (b), and the trimmed trips used for estimation (c). We  observe that the trimmed trips preserve a good coverage of the network.

\subsection{Computing predictions}

The predicted flow for a trip starting in $o \in \mathcal{O}$ and ending in $d \in \mathcal{D}$ is the flow vector $\hat{x}^{od}$ that minimizes the cost

\begin{equation}
C(x) = \sum_{e\in\mathcal{E}} l_e  \left( c_e  x_e + F(x_e) \right), 
 \label{eq:USI}
\end{equation}
subject to the flow conservation constraint. The generalized cost association with this OD pair is $C(\hat x^{od})$.

The edge cost rates $c_e , e \in \mathcal{E}$ are found by multiplying the corresponding row in the  link attribute matrix $Z_e$ with the $\hat{\beta}$ parameters estimates reported in Table \ref{tab:model}, i.e., $c_e = Z_e \hat{\beta}$. 
The minimization problems are solved using conic optimization in the software Mosek Fusion \textcite{mosek}.
We  define the average length between $o$ and $d$ corresponding to the predicted flows as $\hat{l}^{od} = \sum_e \hat{x}^{od}_e l_e$. 
Finally, we obtain the predicted average generalized cost between $o$ and $d$, omitting the perturbation term, as $\hat{c}^{od} = \sum_e c_e \hat{x}^{od}_e$.

\section{Route choice model}
\label{sec:S route choice model}

\subsection{Estimation results}
Table \ref{tab:model} shows the estimated parameters for the preferred model specification, along with clustered heteroscedasticity consistent standard errors. To aid interpretation, the last column of the table shows the parameters divided by the parameter for the constant; thus, the scaled parameters express the subjective cost rate in terms of metres traveled on the reference category road. We discuss the results in terms of the scaled parameters.

The reference category is residential roads without specific bicycle infrastructure in low-rise urban areas. The parameter for the constant thus represents the subjective cost of traveling 1 m by bicycle on the reference category. 

The next set of parameters measures the impact of various mutually exclusive infrastructure types on the link cost rate. 
Links with stairs (intended for pedestrians) are classified as a separate category and incur a penalty of 76\%. 
The cost rate is up to 27\% higher for infrastructure types that are shared with pedestrians. The additional cost rate for ``living streets'' is not significantly different from zero.\footnote{ Small residential streets with parked cars and no bicycle infrastructure, where there may be children playing etc.}

Compared with the reference, bicyclists have some preference against large roads (roads with at least two lanes in one direction, 11\%), whereas the preference against medium roads (roads with at most one lane in each direction) is small and statistically insignificant.

Provision of dedicated bicycle infrastructure quite substantially reduces the subjective cost of bicycling. Cycleways (bicycle paths in own trace) have 20\% lower cost rate than the reference. On residential and medium roads, bicycle lanes, whether protected or just painted, reduce the cost rate by 14\% and 22\%, respectively. The type of bicycle lane has a significant effect on the cost rate for the large roads category: painted bicycle lanes have only a small and statistically insignificant effect on cost rate, whereas protected bicycle lanes reduce the cost rate by 34\%. It makes clear intuitive sense that the impact of bicycle lanes is larger the larger the road, and that only protected lanes affect the largest roads where car traffic is heavier.

Provision of bicycle-friendly infrastructure thus has a substantial effect on route choice. As evident from Section ``Bicycle travel demand''in the main text, this translates into a substantial effect on the volume of bicycle trips.

A number of routes are marketed as so-called cycle superhighways. This label is applied to high-quality, continuous bicycle routes built to cater to commuter cyclists.
The cost rate on the links of these routes is 12\% lower than that on similar links without the cycle superhighway label. The cycle superhighway label and the associated infrastructural changes are likely the cause of the reduction in subjective cost. However, the cycle superhighway label likely has just been attached to routes that were already attractive. To check this, we include a variable indicating routes that are proposed to become cycle superhighways in the future. We find that the cost reduction associated with these links is almost exactly the same as the cost reduction found for the actual cycle superhighway links. The model already accounts for a range of link characteristics, including upgrades to the bicycle infrastructure that take place in the process of creating a cycle superhighway. The attraction of the actual and proposed cycle superhighways could therefore be attributable to route-level and not link-level features; perhaps the feature that these routes are high-quality, continuous bicycle routes \parencite{Heinen2009,liu2019practitioners}. The similarity of the parameters suggests that not the labeling but rather the fact that already attractive routes have been selected to receive the cycle superhighway label makes the difference. 

The next set of parameters accounts for the land use near cycleways. We treat cycleways separately, as they turned out to act different from the other infrastructure types.
The cost rate is much reduced for cycleways in industrial areas (48\%) or green areas (53\%) compared to low-rise urban areas. It makes intuitive sense that cycleways in green areas may be pleasant. Another potential explanation that also applies to industrial areas is the attractiveness of isolation from heavy traffic. The parameters for cycleways near water or open landscape are not statistically significant.

For the other infrastructure types, the subjective cost is lower near all other land uses than low-rise urban areas, but the differences are not statistically significant for all land use types. The largest cost reduction is found for links near water (26\%).

The last set of parameters concerns some special link characteristics. 
The elevation gain variable measures the total elevation gain on a link that has a gradient of 3.5 percent or more. It aggregates the vertical distance on the parts of the links where the slope is at least +3.5\%. The scaled parameter is estimated to be $16.6$, which means that an elevation gain of 0.05 me per meter implies an increase of 83\% in the cost rate, which seems reasonable.  
If the surface is gravel, the cost rate increases by 19\%, whereas the parameter for cobblestones is small and not statistically significant.
Finally, going against (car) traffic on one-way streets increases the cost rate by 59\%.

\subsection{Model validation} \label{sec:Validation}

\subsubsection{Comparison of observed and predicted flows by infrastructure type}

Using the estimated parameters, we compute the predicted flows for each OD pair. Table \ref{tab:usage} compares the observed and predicted flows, showing the percentage of flow that occurs on each link type. For comparison, the first column of the table shows the corresponding shares weighted just by link lengths, which is what would be the result if traffic was distributed on the network at random. We find, supporting the model, that the observed and predicted flow shares are very similar. They are both very different from the network shares, especially because cyclists use roads with bicycle infrastructure much more than would have been the case if traffic was randomly distributed on the network.

\subsubsection{Comparison of observed and predicted flows by OD pair} 

To compare the observed and predicted flow vectors for a given OD pair, we introduce a correlation measure as follows:  defined for convenience as
\[\mathbb{E}_l (x)= \sum_{e \in \mathcal{E}}\frac{l_e}{\sum_{e' \in \mathcal{E}} l_{e'}}   x_e.\] 
We then compute the correlation between the observed and predicted flow on a random kilometer of road (where $\circ$ is the Hadamard product):

\begin{equation}\rho^{od} = 
 \rho( x^{od};\hat x^{od})=  
 \frac{\mathbb{E}_l (x^{od} \circ  \hat x^{od}) -\mathbb{E}_l (x^{od}) \mathbb{E}_l (\hat x^{od})}
 {\sqrt{ \mathbb{E}_l (x^{od}\circ x^{od})  - (\mathbb{E}_l (x^{od}))^2} \sqrt{ \mathbb{E}_l (\hat x^{od}\circ \hat x^{od})  - (\mathbb{E}_l (\hat x^{od}))^2}}
 . \label{eq:ValidationCorrelation}
\end{equation}
The correlation is bounded between $-1$ and $1$, with $1$ corresponding to perfect correlation.

\begin{figure}[H]
    \begin{subfigure}{0.49\textwidth}
    \includegraphics[width=\textwidth]{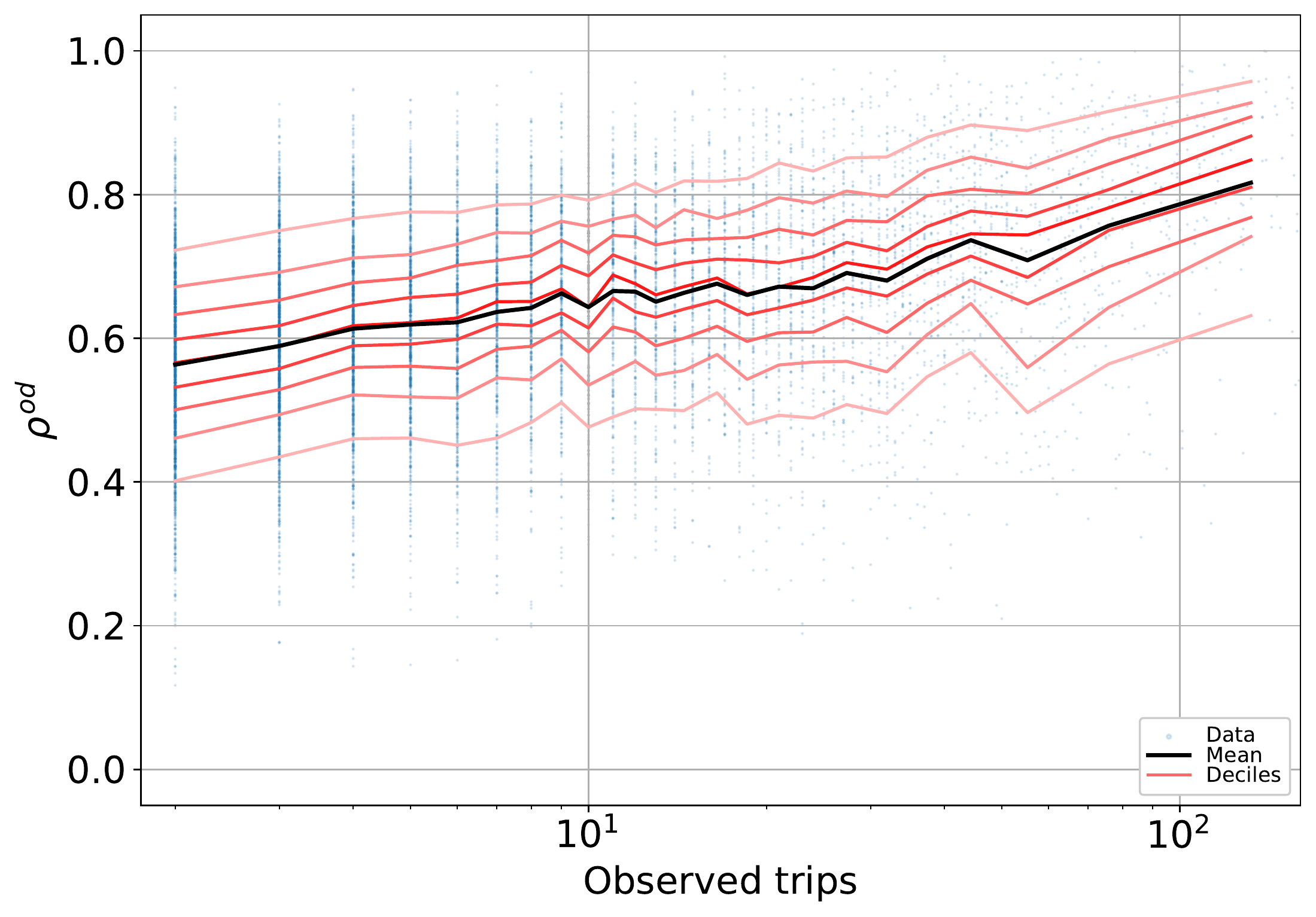}
    \caption{Number of observed trips}
    \end{subfigure}
    \hfill
    \begin{subfigure}{0.49\textwidth}
    \includegraphics[width=\textwidth]{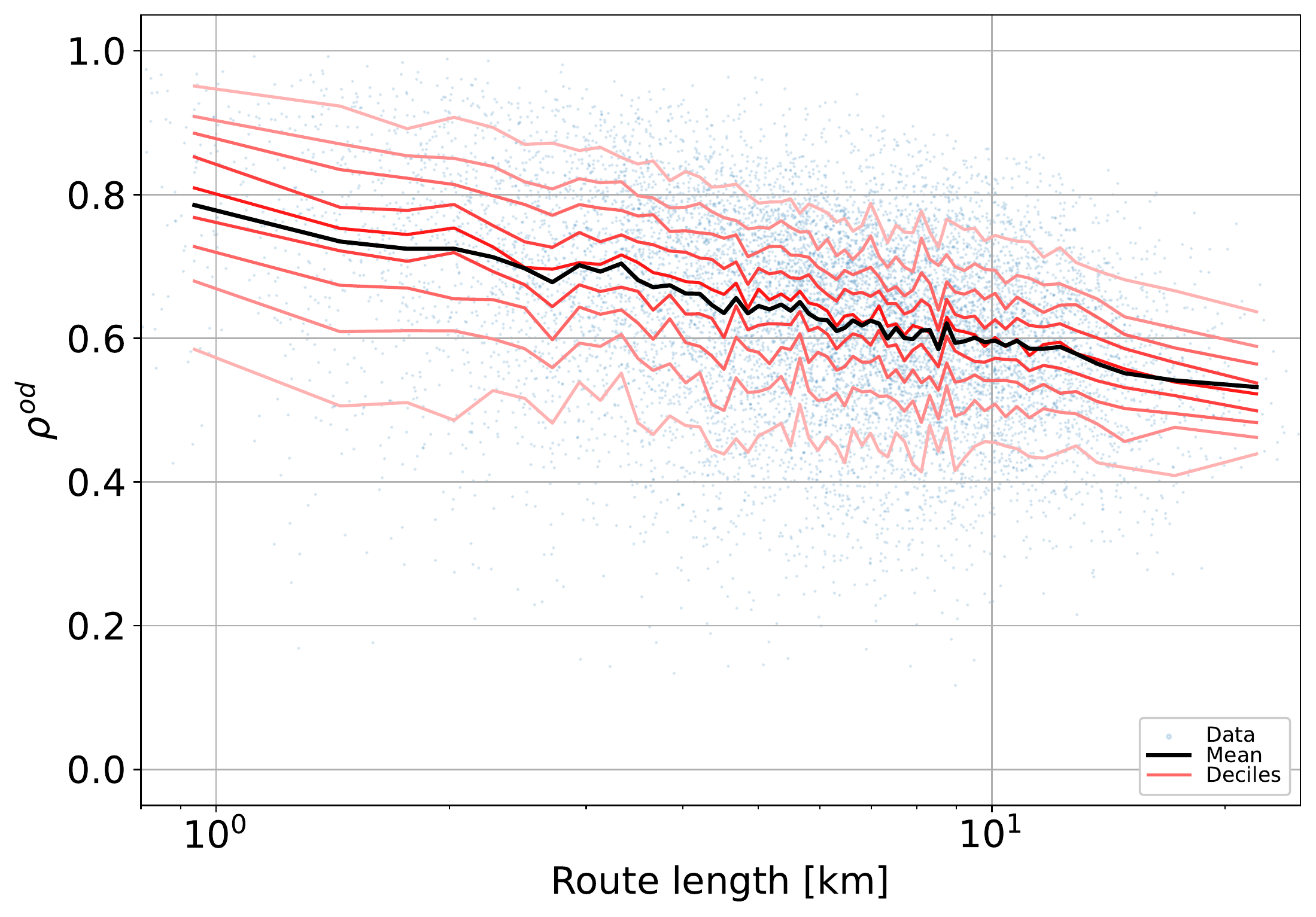}
    \caption{Expected route length}
    \end{subfigure} 
    \caption{Correlation between predicted and observed flow as defined in \eqref{eq:ValidationCorrelation} plotted against the number of observed trips per OD pair (panel a) and the expected route length per OD pair (panel b). The black line indicates mean correlation, and the red lines indicate the corresponding binned deciles. }
    \label{fig:ValidationPlotCorrelation}
\end{figure}

We compute the correlation for every OD pair in the data. Figure \ref{fig:ValidationPlotCorrelation} plots the correlation against the number of observed trips in each OD pair and against the expected route length in each OD pair. We find that the mean correlation ranges from more than $0.5$ to more than $0.75$. The correlation increases with the increase in the number of observed trips, which is reasonable since sampling noise causes the observed flow to differ from the predicted flow and hence decreases the correlation. The correlation decreases with route length, which may also be attributable to sampling noise because the number of observed trips decreases with trip length. 

The average correlation across all the OD pairs used for the route choice estimation is found to be $0.6356$. When weighting by the number of observed trips, to obtain the correlation between the observed and predicted link flows for an average observed trip, the correlation is $0.7001$.

\subsubsection{Changing the number of ODs}\label{sec:ChangingNumberOfODs}
As mentioned in Section \ref{Sec:SI data and data processing}\ref{sec:Data:OD}, the number of origins and destinations needs to be chosen prior to the estimation. As a robustness check, we estimate the model using 100 and 400 Os and Ds in addition to the 200 used for the main result. Table \ref{tab:model100200400} shows the estimates, which are broadly similar across models.

\subsubsection{Cross-validation} \label{sec:CrossValidation}
Because we use a very large dataset with more than 1.8 million rows in the linear regression equation, our model is not very likely to suffer from over-fitting. This expectation is confirmed by a test where we randomly split the OD pairs into two separate datasets (data split A and data split B). Data split A comprises 5,520 OD pairs with  929,569 rows for the linear regression, and data split B comprises 5,519 OD pairs with 937,198 rows for the linear regression.
Table \ref{tab:modelAFullB} shows the parameter estimates from the route choice model for the two data splits A and B, with the estimates from the model based on the full sample included for easy reference. The parameter estimates appear to be quite stable across splits. 

\begin{figure}[htb]
\begin{subfigure}{0.32 \textwidth}
    \includegraphics[width=\textwidth]{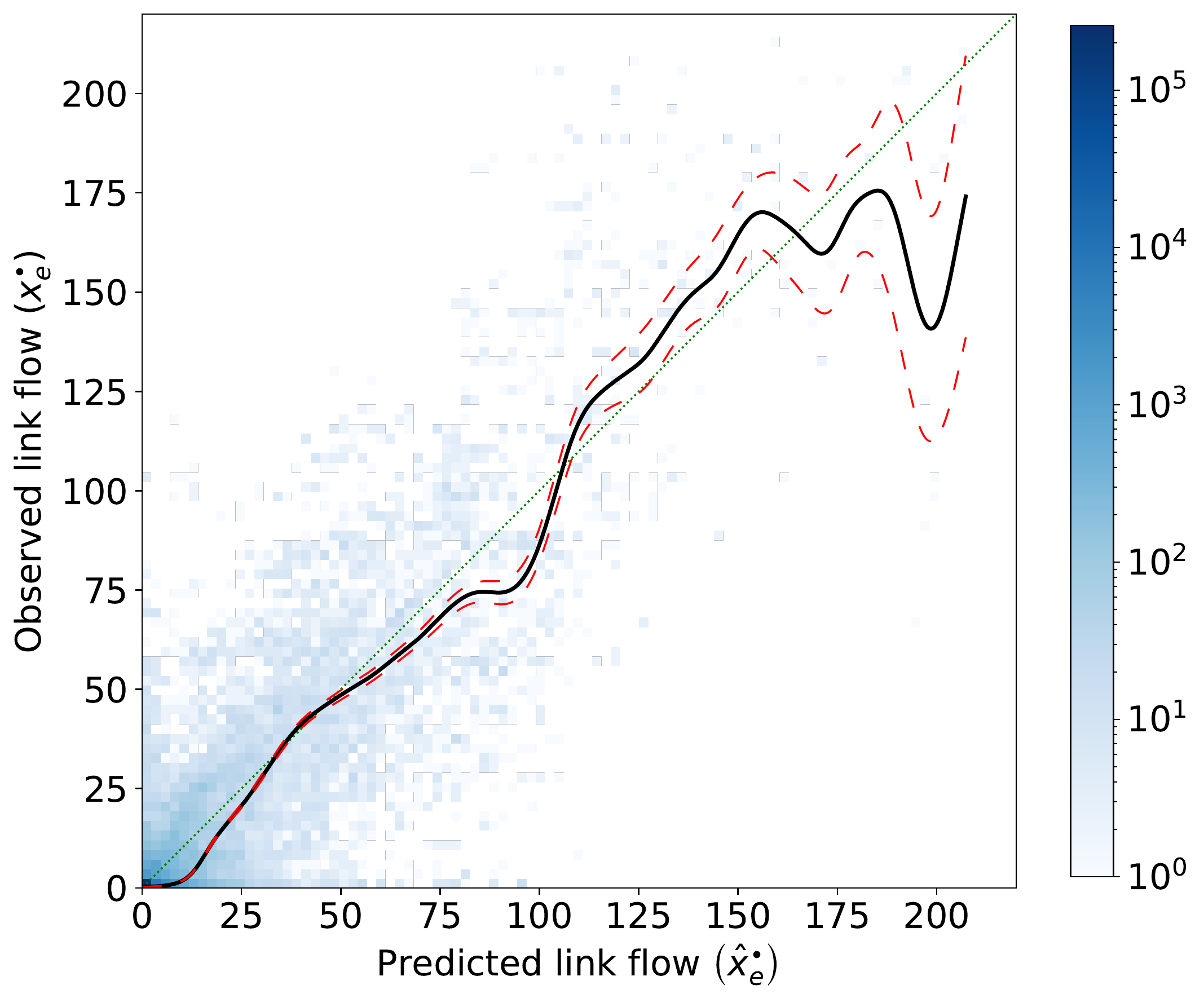}
    \caption{Model based on data split A} \label{fig:ValidationPartCStyle_AFullB_A}
\end{subfigure} \hfill
\begin{subfigure}{0.32 \textwidth}
    \includegraphics[width=\textwidth]{FiguresPaper/Validation/TotalLinkFlowValidation.pdf}
    \caption{Model based on the full data} \label{fig:ValidationPartCStyle_AFullB_Full}
\end{subfigure} \hfill
\begin{subfigure}{0.32 \textwidth}
    \includegraphics[width=\textwidth]{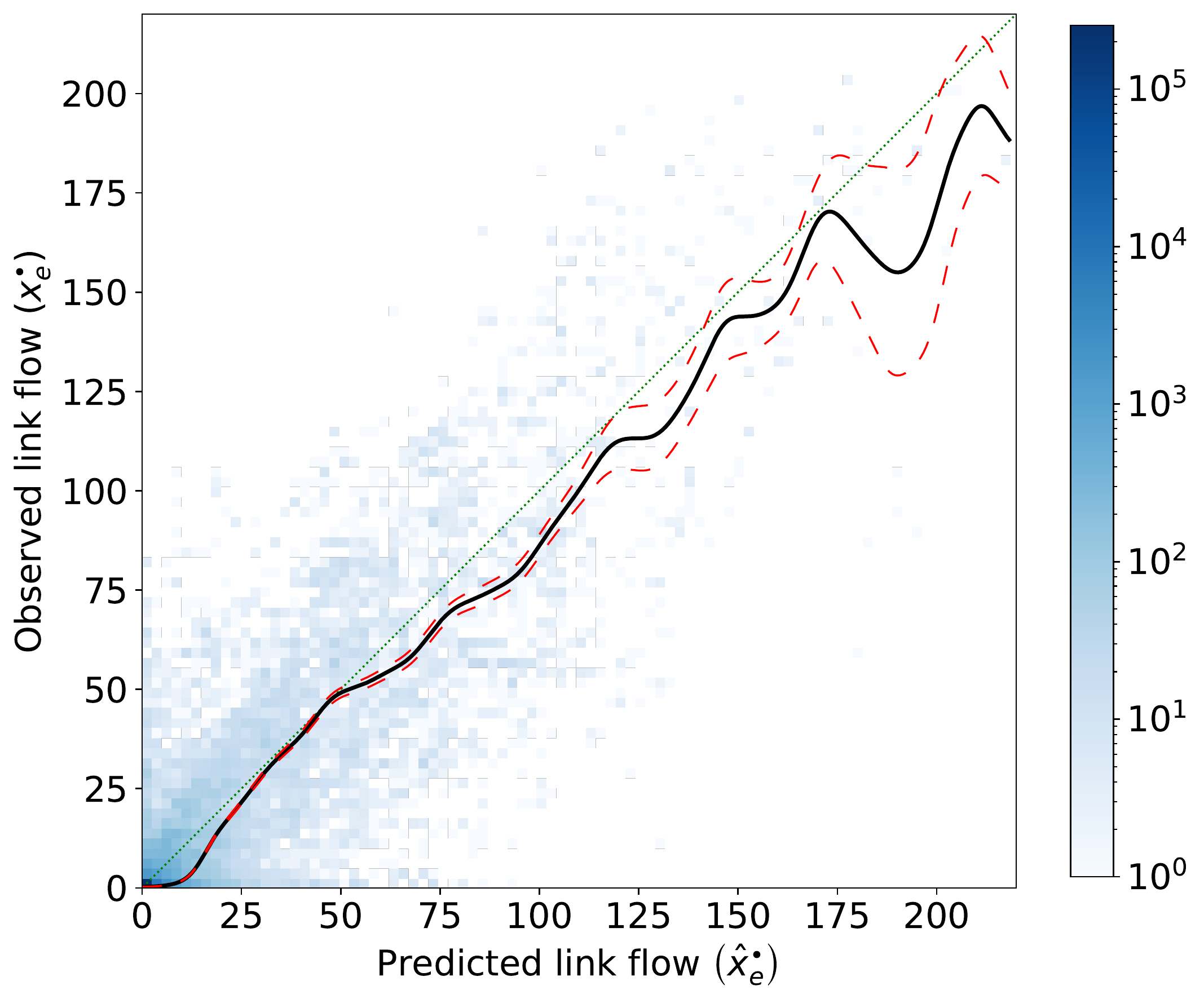}
    \caption{Model based on data split B} \label{fig:ValidationPartCStyle_AFullB_B}
\end{subfigure}
   
    \caption{Heatmap of total observed link flow $\left(x_e^{\bm\cdot} = \sum_{o\in \mathcal{O}} \sum_{d \in \mathcal{D}} x_e^{od}\right)$ against the total predicted link flow $\left(\hat x_e^{\bm\cdot} = \sum_{o\in \mathcal{O}} \sum_{d \in \mathcal{D}} \hat x_e^{od}\right)$ for each link $e \in \mathcal{E}$. The color of each grid cell represents the number of links belonging to that cell. The thin green dotted line is the $45^\circ$ line.
    The black line is a Nadaraya-Watson non-parametric regression \parencite{Nadaraya1964,Watson1964} with Gaussian kernel and bandwidth chosen by eyeballing (a bandwidth of 5 for data splits A and B, and a bandwidth of 10 for the full data). The corresponding 95\% pointwise confidence band is indicated by dashed red lines. 
    The model based on data split A has been applied out-of-sample on data split B and vice versa.
    }
    \label{fig:ValidationPartCStyle_AFullB}
\end{figure}

As an out-of-sample prediction test, we compute the predicted flows for each sample split, based on the parameters estimated for the other sample split.  Figure \ref{fig:ValidationPartCStyle_AFullB} plots the observed flows vs. the predicted flows for each split combination as well as for the full sample. We find that the out-of-sample fit is very similar to the in-sample fit. The correlations between the observed and out-of-sample predicted  link flows, computed according to Eq. \ref{eq:ValidationCorrelation}, are $0.879$ and $0.889$ for the models based on data split A and data split B, respectively. This is comparable to the correlation of $0.889$ found for the full sample model.  

We similarly calculate correlations between the observed and out-of-sample predicted flows for each OD pair.  The average OD link flow correlation across OD pairs is found to be $0.629$ and $0.638$ ($0.691$ and $0.705$ when weighting by number of observed trips), respectively, when the models based on data split A and data split B are used out-of-sample. This is very similar to the correlation value of $0.634$ ($0.699$ when weighted by number of observed trips) found for the overall model applied in-sample.

\section{Gravity model}
\subsection{Specification of the demand function}
The gravity model assumes that the demand is driven by the cost $\hat c ^{od}$ from the route choice model and not just by the OD distance. To test whether this is the case, we split the cost into a length component and a residual quality component that is orthogonal to length. In this decomposition, we omit the perturbation term, as this term is expected to increase with a decrease in the OD distance because shorter trips have fewer relevant routes. For the test, we therefore compute the length component as the predicted value in a linear regression of the cost $\hat c ^{od}$ while excluding the perturbation term against the predicted trip length. The quality component is the residual from this regression. We have then estimated the gravity model in \eqref{eq:Gravity2} with two demand functions, one for length and one for quality. As before, both are specified as piece-wise linear. 
    \begin{equation}
    \ln E \left[Y^{od}\right] = D_1\left( length^{od}\right) +D_2\left( quality^{od}\right)+ \delta + \eta_o + \gamma_d   
    \label{eq:Gravity2}
    \end{equation}

\begin{figure}[H]
    \centering
    \includegraphics[width=0.6 \textwidth]{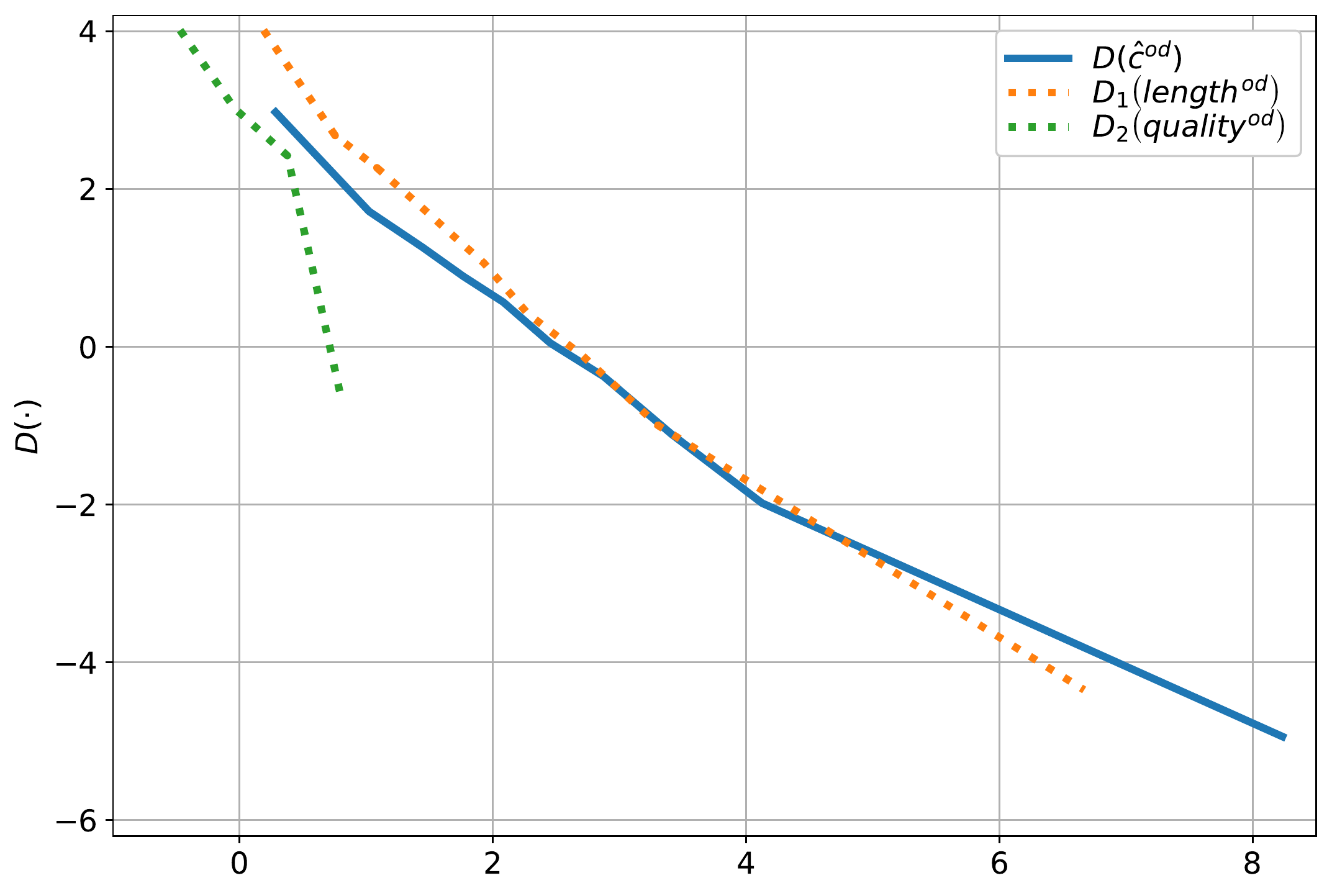}
    \caption{Estimated piecewise linear specifications of $D$ (from \eqref{eq:Gravity} in the main text) for three different explanatory variables. Blue: Demand function $D$ from the original model (\eqref{eq:Gravity} in the main text). Orange \& green: Demand functions $D_1$ and $D_2$ from the model in \eqref{eq:Gravity2}, respectively.
    }
    \label{fig:GravityVariables}
\end{figure}

Figure \ref{fig:GravityVariables} shows in blue the estimated demand function from the original model (\eqref{eq:Gravity} in the main text) along with the demand functions from the model in \eqref{eq:Gravity2}, where the orange dotted curve is the influence of length on demand and the green dotted curve is the influence of quality on demand. We observe that the curve of the influence of quality is least as steep as the curve of the estimated demand function from the original model. We therefore conclude that collapsing the two effects (length and quality) into a single demand function yields a conservative estimate of the effect on demand.

\subsection{Demand elasticity}

\begin{figure}[htb]
    \centering
    \includegraphics[width = 0.7 \textwidth]{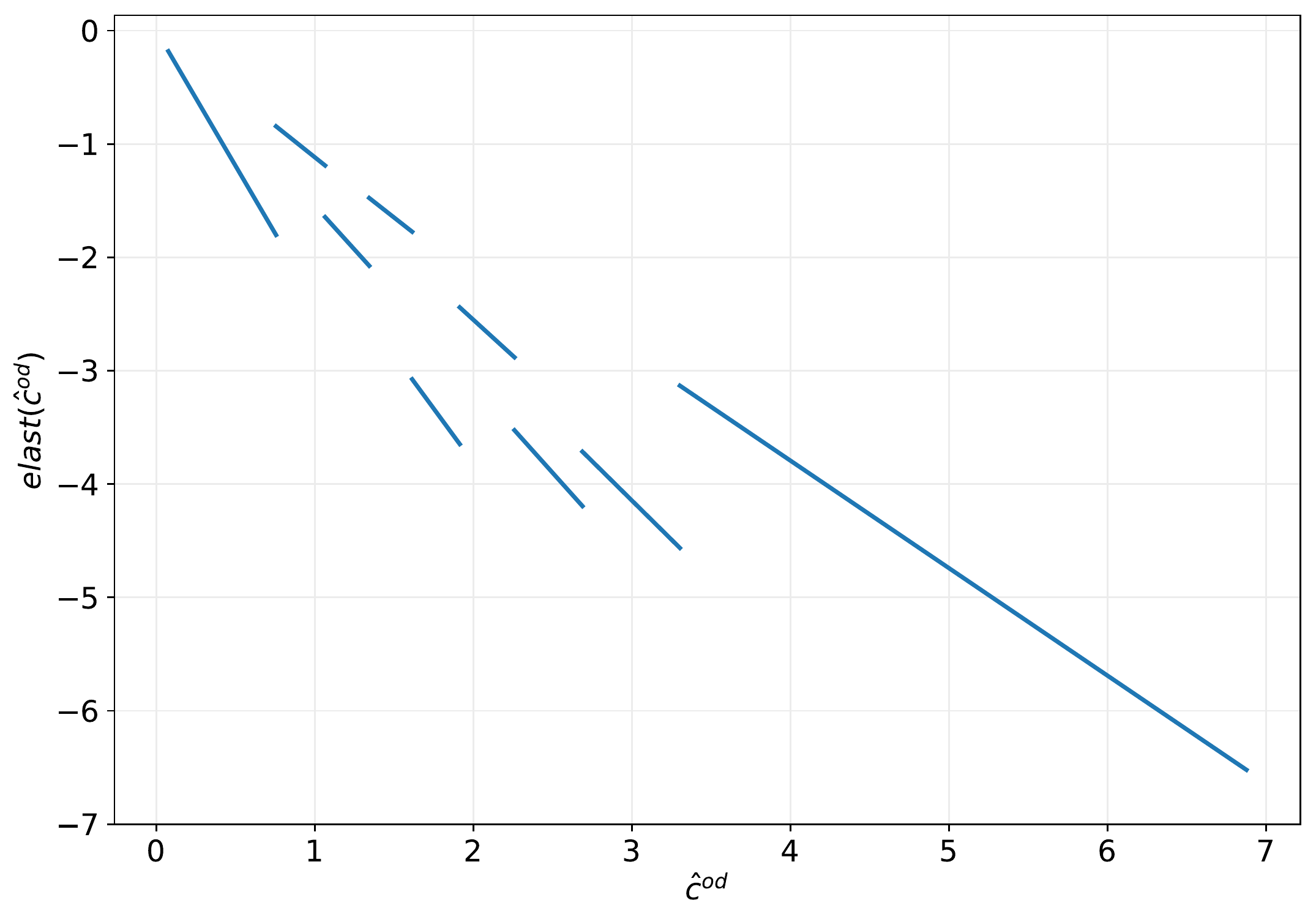}
    \caption{Demand elasticities ($elast(\hat c ^{od})$) calculated from the estimated demand function using $elast(\hat c ^{od}) = D'(\hat c ^{od}) \hat c ^{od}$. The demand elasticity at a point $\hat c ^{od}$ is the relative change in demand per relative change in cost. }
    \label{fig:DemandElasticities}
\end{figure}

The elasticity of demand is the relative change in demand per relative change in cost. Thus, an elasticity of -1 implies that a 10\% increase in cost leads to a 10\% decrease in demand. Figure \ref{fig:DemandElasticities} shows the elasticities calculated along the estimated demand curve. The result indicates that the elasticity decreases about monotonically with the cost.

\section{Counterfactuals}\label{sec:CountF}

\subsection{Method} \label{sec:CountF:Method}

Using the average predicted generalized OD costs $\hat{c}^{od}$, the estimated function $D$, and values of $\delta$, $\eta_o$, and $\gamma_d$, we apply the estimated gravity model to compute $\hat{Y}^{od}$, the predicted number of trips between any $o \in \mathcal{O}$ and $d \in \mathcal{D}$. 
By weighting each OD flow prediction with the corresponding predicted average length between $o$ and $d$, we find the total sum of predicted kilometers traveled in our sample. 
This number corresponds to the size of our sample of bicycle trips. We scale it to the actual annual number of kilometers traveled by bicycle in the study area. We can estimate this roughly to be 1,026 million km, computed using \textcite{RegionHovedstaden2020} as the total kilometers in Frederiksberg and Copenhagen municipalities, and 50\% of the kilometers in the suburbs. In this way, we find that each predicted trip represents  $\zeta=\text{1,943}$ annual trips.

\begin{figure}[htb]
    \centering
    \begin{subfigure}[t]{0.32\textwidth}
  \centering
  \includegraphics[width=\textwidth]{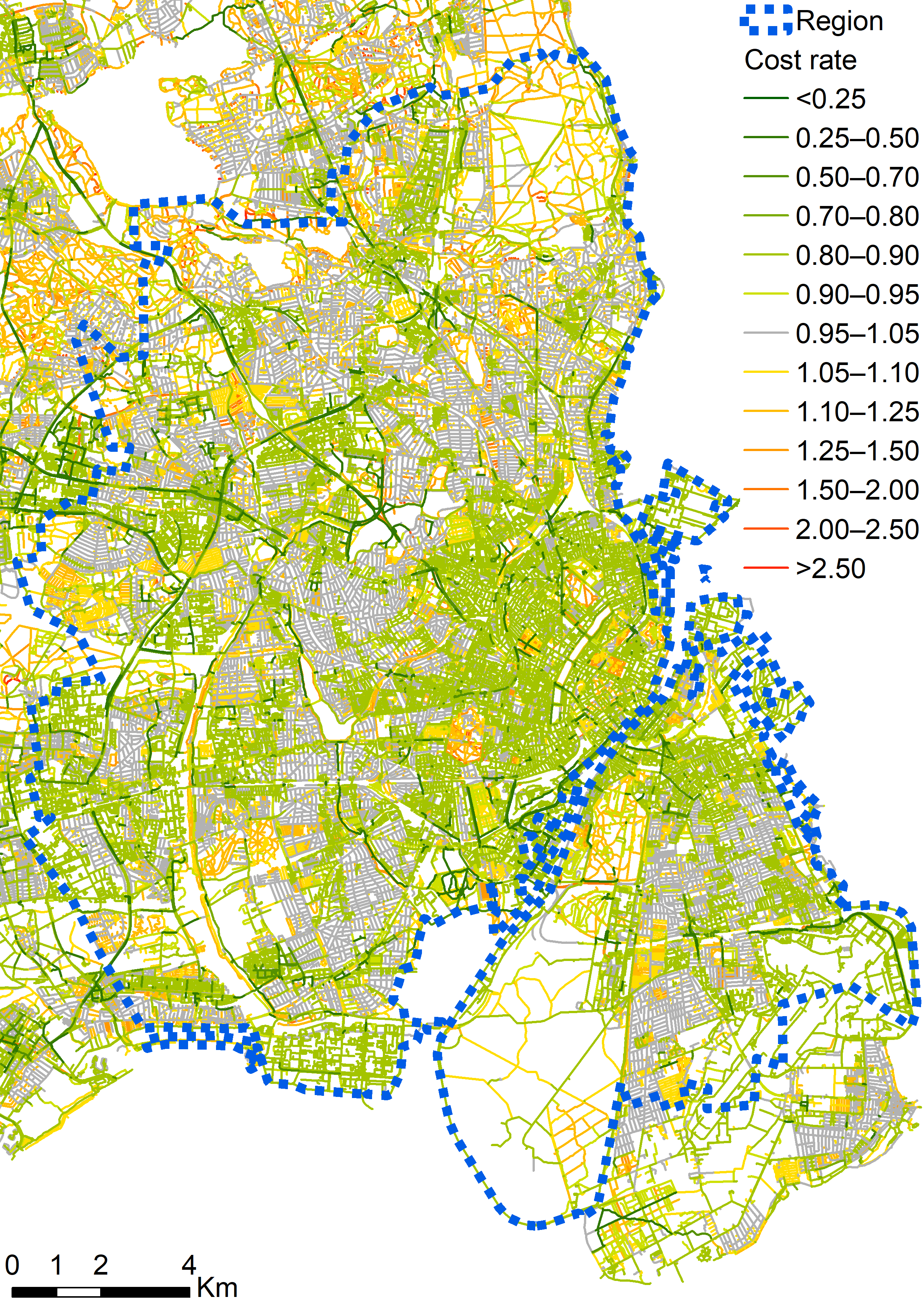}
  \caption{No bicycle lanes}
  \label{fig:CounterfactualsNoPLSC}
   \end{subfigure}%
      \hfill
     \begin{subfigure}[t]{0.32\textwidth}
  \centering
  \includegraphics[width=\textwidth]{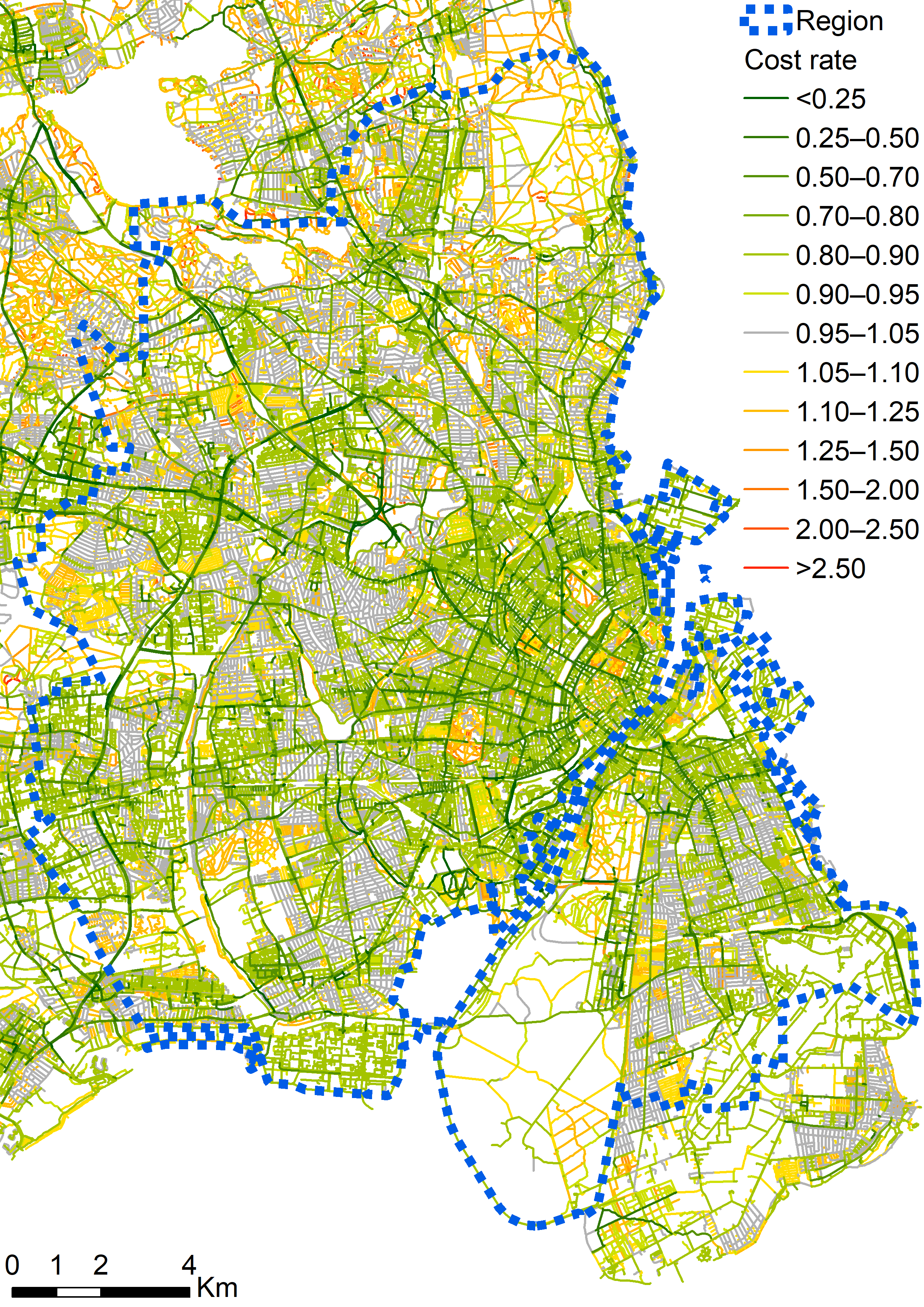}
  \caption{Painted lanes instead of protected on large roads}
  \label{fig:CounterfactualsLPToll}
    \end{subfigure}%
    \hfill
    \begin{subfigure}[t]{0.32\textwidth}
 \centering
  \includegraphics[width=\textwidth]{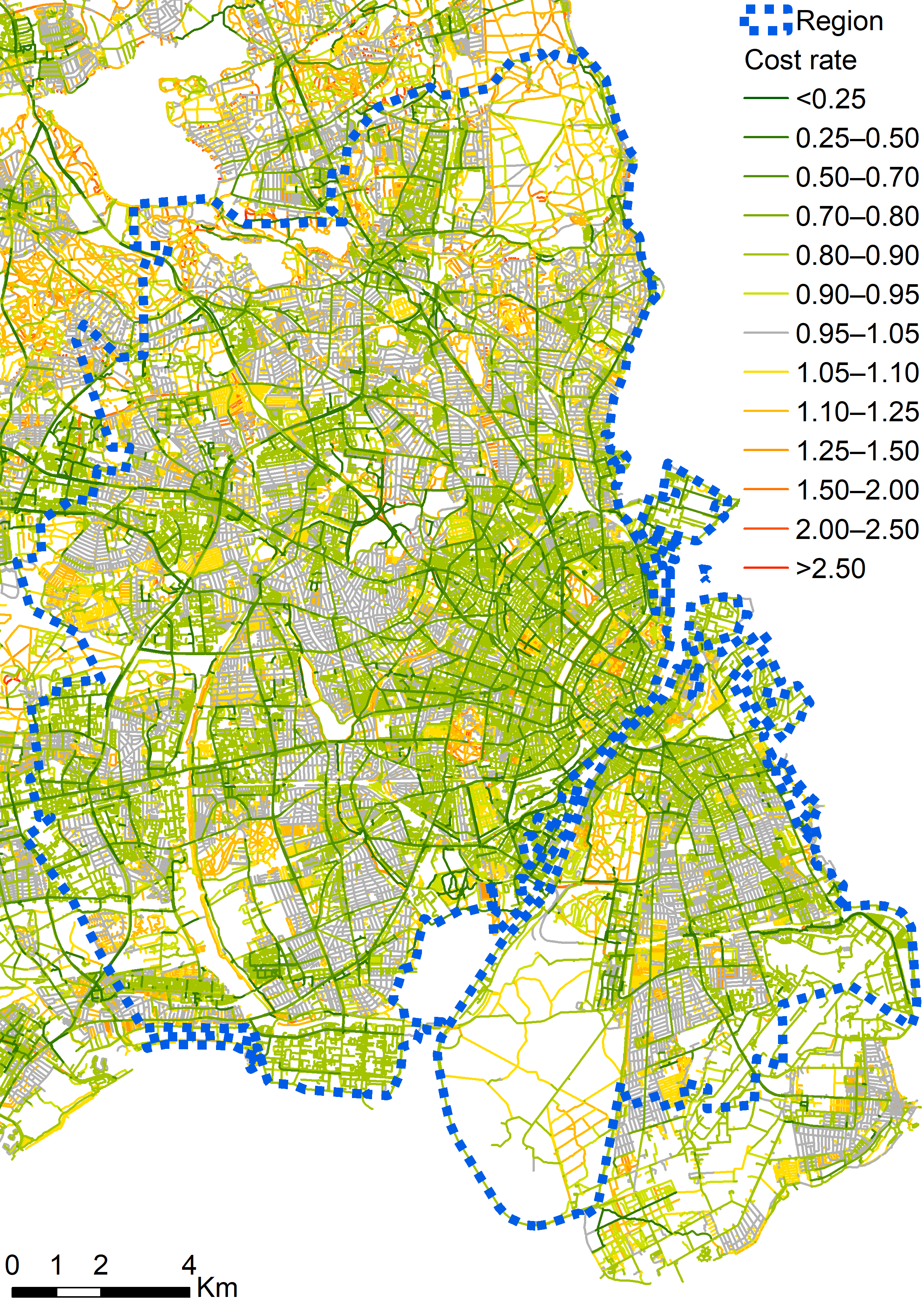}
  \caption{No existing or planned cycle superhighway classifications}
  \label{fig:CounterfactualsNoSC}
    \end{subfigure}%
    \caption{The estimated cost rates of links in the three counterfactual scenarios. The gray links mark the reference case: residential roads with no bicycle infrastructure in low-rise urban areas, scaled to a value of 1. Shades of green correspond to increasingly more attractive links, and shades of orange and red correspond to increasingly less attractive links.}
    \label{fig:Counterfactuals}
\end{figure}

When simulating a counterfactual $s$, we first adjust the link attribute matrix $Z$ according to the counterfactual scenario and then obtain a new link attribute matrix $Z^s$, and corresponding link-specific costs rates $\tilde{c}_e^s = Z_e^s \hat{\beta}$ (visualized for our three counterfactuals in Figure \ref{fig:Counterfactuals}). Using these new link cost rates, we recompute the cost minimizing flows, which we denote as $\hat{x}^{od,s}$. We also compute the average predicted generalized OD costs $\hat{c}^{od,s} = \sum_{e \in \mathcal{E}} l_e \hat{\tilde{x}}_e^{od,s} c_e^s$. We keep the estimated  $\delta$, $\eta_o$, and $\gamma_d$ from the base scenario, and scale each trip with the same factor $\zeta$ as in the base scenario. 

The relative cost increase in  counterfactual $s$ compared with the base scenario is

\begin{equation}
\frac{
\sum\limits_{o \in \mathcal{O}} \sum\limits_{d \in \mathcal{D}} 
 \hat{Y}^{od}\frac{\hat{c}^{od,s}-\hat{c}^{od}}{\hat{c}^{od}}}{\sum\limits_{o \in \mathcal{O}} \sum\limits_{d \in \mathcal{D}} 
\hat{Y}^{od}} . \label{eq:CostIncrease2}
\end{equation}

The relative decrease in the number of trips in counterfactual $s$ relative to the base scenario is

\begin{equation}
1 - 
     \frac{\sum\limits_{o \in \mathcal{O}}\sum\limits_{d \in \mathcal{D}}  \hat{Y}^{od,s}} {\sum\limits_{o \in \mathcal{O}}\sum\limits_{d \in \mathcal{D}} \hat{Y}^{od}} \label{eq:TripsDecrease}.
\end{equation}

The corresponding measure for the number of kilometers traveled is

\begin{equation} 
1 - 
     \frac{\sum\limits_{o \in \mathcal{O}}\sum\limits_{d \in \mathcal{D}} \hat{Y}^{od,s} \hat{l}^{od,s}} {\sum\limits_{o \in \mathcal{O}}\sum\limits_{d \in \mathcal{D}} \hat{Y}^{od}\hat{l}^{od}}. \label{eq:TripsKilometres}
\end{equation}

To compute the consumer surplus \textcite{Fosgerau2021b}, we first convert the generalized cost to length units by dividing the it with the constant term in $\hat{\beta}$ (0.4555). 
Next, we use the average speed in our sample (14.84 km/h) to convert the length units into travel time units. The change in travel time then becomes

\begin{equation}
    \Delta T^{od,s} = \frac{1}{14.84}\frac{\hat{c}^{od,s}-\hat{c}^{od}}{0.4555}.
\end{equation}

The value of time for cyclists in Denmark is 16.13 € per hour according to the official guidelines \textcite{Enhedspriser2022}. Combining the change in travel time, the value of time, and the change in demand, we  apply the rule of a half to compute the change in consumer surplus for each counterfactual scenario:

\begin{equation}
     16.13 \cdot \zeta \cdot   \sum_{o \in \mathcal{O}} \sum_{d \in \mathcal{D}} \Delta T^{od,s} \frac{\hat{Y}^{od} + \hat{Y}^{od,s}}{2}.
\end{equation}

Health and accident benefits are computed multiplying the combined unit price of 0.91 €  (1.17 € for health and -0.26 € for accidents \textcite{Enhedspriser2022}) by the scaled change in kilometers traveled to get the monetary annual benefit

\begin{equation}
    0.91 \cdot \zeta \cdot \sum_{o \in \mathcal{O}} \sum_{d \in \mathcal{D}} \hat{Y}^{od,s} \hat{l}^{od,s}  - \hat{Y}^{od}\hat{l}^{od}.
\end{equation}

The societal benefit owing to cyclists' travel time, health, and accidents is the sum of the  consumer surplus benefits and the health/accident benefits. 
The societal benefit per changed network length is determined by dividing the societal benefit with the sum of the (directed) link lengths of the links that were changed to create the counterfactual $s$.  However, for the third counterfactual on cycle superhighways, the corresponding route distances were used instead of the network length, as the expense related to this counterfactual occurs at the route level.

\newpage

\newpage

\newpage

\newpage

\newpage

\newpage

\newpage

\begin{table}[H] 
\footnotesize
\centering
\hspace*{-1cm}
\resizebox{\textwidth}{!}{\begin{tabular}{ll}  \toprule
\textbf{Infrastructure type} & \textbf{Description} \\ \midrule
 Residential roads w/o bicycle infrastructure & \multirow{3}{0.57 \textwidth}{Residential roads: Roads with OSM tag \textit{highway=`residential'}}
 \\
 Residential roads w/ painted bicycle lanes & \\
 Residential roads w/ protected bicycle lanes & \\[0.5em]
 Medium roads w/o bicycle infrastructure &  \multirow{3}{0.57\textwidth}{Medium roads: Roads with OSM tags \textit{highway=\{`primary',}\textit{`secondary',`tertiary',`unclassified'\}},  that have at most one car lane per direction.}\\
 Medium roads w/ painted bicycle lanes & \\
 Medium roads w/ protected bicycle lanes & \\[0.5em]
 Large roads w/o bicycle infrastructure&  \multirow{3}{0.57 \textwidth}{Large roads: Roads with OSM tags \textit{highway=\{`primary',}\textit{`secondary',`tertiary',`unclassified'\}}, that have at least two car lanes in at least one direction.}\\
 Large roads w/ painted bicycle lanes & \\
 Large roads w/ protected bicycle lanes & \\[0.5em]
 Cycleways & OSM tag \textit{highway=`cycleway'}\\[0.5em]
 Footways & OSM tag  \textit{highway=`footway'}\\[0.5em]
 Living streets & OSM tag \textit{highway=`living\_street'} \\[0.5em]
 Shared paths & OSM tags \textit{highway=\{`path',`track',`service'}\}\\[0.5em]
 Pedestrian zones & OSM tag \textit{highway=`pedestrian'}\\[0.5em]
 Stairs & OSM tag \textit{highway=`steps'}\\ \bottomrule
\end{tabular}}
 \caption{Network attributes related to infrastructure type and their associated OSM tags}
 \label{tab:attr}
\end{table}

\newpage
\begin{table}[h]
    \centering
    \resizebox{\textwidth}{!}{\begin{tabular}{l|rrr} \toprule
  & \textbf{Trips} & \textbf{Users} & \textbf{Kilometres}  \\  \midrule
  After initial data filtration & 218,489 & 8,588 & 762,791.8 \\
  Trips connecting candidate $O$s and $D$s & 152,323 & 7,672 & 614,909.7 \\
  Trimmed trips used for estimation & 152,323 & 7,672 & 417,358.0  \\ \bottomrule
    \end{tabular}}
        \caption{Summary of the dataset size after different filtration subprocesses. The subprocesses correspond to panels (a), (b), and (c) in Figure \ref{fig:GPStrim}, respectively.}
    \label{tab:DataFiltration}
\end{table}

\newpage

\begin{table}[htb] 
\footnotesize
\centering  
\resizebox{\textwidth}{!}{\begin{tabular}{l|rrrr} \toprule
 
\textbf{Parameter} & \textbf{Coef.} & \textbf{Std. err.} & \textbf{\textit{P}-val.} &  \textbf{Scaled} \\
\midrule
\textbf{Constant}  &$-\text{0.456}$&$\text{0.027}$&***&$\text{1}$\\
\textbf{Infrastructure}  &&&&\\
\tabindent Stairs  &$-\text{0.342}$&$\text{0.096}$&***&$\text{0.750}$\\
\tabindent Pedestrian zones  &$-\text{0.121}$&$\text{0.025}$&***&$\text{0.265}$\\
\tabindent Footways  &$-\text{0.090}$&$\text{0.012}$&***&$\text{0.198}$\\
\tabindent Shared paths  &$-\text{0.026}$&$\text{0.016}$&&$\text{0.057}$\\
\tabindent Living streets  &$-\text{0.051}$&$\text{0.039}$&&$\text{0.113}$\\
\tabindent Cycleways  &$\text{0.089}$&$\text{0.037}$&*&$-\text{0.195}$\\
\tabindent {Residential roads}   &&&&\\
\tabindent \tabindent {No bicycle infrastructure}  & ------- & ------- & ------- & ------- \\
\tabindent \tabindent W/ bicycle infrastructure  &$\text{0.065}$&$\text{0.019}$&***&$-\text{0.142}$\\
\tabindent Medium roads  &$-\text{0.005}$&$\text{0.018}$&&$\text{0.010}$\\
\tabindent \tabindent {No bicycle infrastructure}  & ------- & ------- & ------- & ------- \\
\tabindent \tabindent W/ bicycle infrastructure  &$\text{0.101}$&$\text{0.018}$&***&$-\text{0.221}$\\
\tabindent Large roads  &$-\text{0.050}$&$\text{0.015}$&***&$\text{0.110}$\\
\tabindent \tabindent {No bicycle infrastructure}  & ------- & ------- & ------- & ------- \\
\tabindent \tabindent W/ painted bicycle lanes  &$\text{0.014}$&$\text{0.024}$&&$-\text{0.030}$\\
\tabindent \tabindent W/ protected bicycle lanes  &$\text{0.154}$&$\text{0.016}$&***&$-\text{0.338}$\\
\textbf{Bicycle route classification}  &&&&\\
\tabindent {No classification}  & ------- & ------- & ------- & ------- \\
\tabindent Cycle superhighway  &$\text{0.057}$&$\text{0.014}$&***&$-\text{0.126}$\\
\tabindent Proposed cycle superhighway  &$\text{0.055}$&$\text{0.012}$&***&$-\text{0.121}$\\
\textbf{Land use, cycleways} &&&&\\
\tabindent High-rise urban areas  &$\text{0.119}$&$\text{0.041}$&**&$-\text{0.260}$\\
\tabindent {Low-rise urban areas}  & ------- & ------- & ------- & ------- \\
\tabindent Industrial areas  &$\text{0.221}$&$\text{0.052}$&***&$-\text{0.486}$\\
\tabindent Green areas  &$\text{0.243}$&$\text{0.054}$&***&$-\text{0.534}$\\
\tabindent Areas near water  &$\text{0.043}$&$\text{0.052}$&&$-\text{0.093}$\\
\tabindent Open landscape  &$-\text{0.037}$&$\text{0.123}$&&$\text{0.082}$\\
\textbf{Land use, other infrastructure} &&&&\\
\tabindent High-rise urban areas  &$\text{0.088}$&$\text{0.024}$&***&$-\text{0.193}$\\
\tabindent {Low-rise urban areas}  & ------- & ------- & ------- & ------- \\
\tabindent Industrial areas  &$\text{0.072}$&$\text{0.038}$&&$-\text{0.158}$\\
\tabindent Green areas  &$\text{0.054}$&$\text{0.037}$&&$-\text{0.119}$\\
\tabindent Areas near water  &$\text{0.118}$&$\text{0.030}$&***&$-\text{0.259}$\\
\tabindent Open landscape  &$\text{0.083}$&$\text{0.096}$&&$-\text{0.182}$\\
\textbf{Elevation gain, $>\,$35 m/km}  &$-\text{7.559}$&$\text{3.682}$&*&$\text{16.595}$\\
\textbf{Surface type}  &&&&\\
\tabindent {Asphalt}  & ------- & ------- & ------- & ------- \\
\tabindent Cobblestones  &$-\text{0.014}$&$\text{0.019}$&&$\text{0.030}$\\
\tabindent Gravel  &$-\text{0.086}$&$\text{0.015}$&***&$\text{0.189}$\\
\textbf{Wrong way}  &$-\text{0.266}$&$\text{0.007}$&***&$\text{0.584}$\\
\midrule
\textbf{\textit{N}} & \multicolumn{4}{c}{1,866,767} \\ 
\bottomrule
\end{tabular}}
 \caption{The estimated parameters for the bicycle route choice model in the main text (200 origins and 200 destinations). Standard errors are clustered (per OD pair) and are heteroscedasticity-consistent. The scaled values are scaled such that the parameter ``Constant'' has a value of 1. Significance levels are *** \textit{P}-value $\leq$ 0.001, ** \textit{P}-value $\leq$ 0.01, and * \textit{P}-value $\leq$ 0.05. \textit{N} denotes the number of rows in the linear regression equation.}
 \label{tab:model}
\end{table}

\newpage

\begin{table}[h] 
\footnotesize
\centering
\resizebox{\textwidth}{!}{\begin{tabular}{l|rrr} 
\toprule &
\textbf{Network [\%]} & \textbf{Observed use [\%]} & \textbf{Predicted use [\%]}\\
\midrule
\textbf{Constant}      & 100.00 &  100.00 &   100.00 \\
\textbf{Infrastructure}& &  &   \\
\tabindent Stairs      &   0.12 &    0.07 &     0.07 \\
\tabindent Pedestrian zones   &   0.19 &    0.72 &     0.35 \\
\tabindent Footways    &   8.24 &    6.80 &     2.49 \\
\tabindent Shared paths&  38.54 &    6.21 &     6.26 \\
\tabindent Living streets     &   0.55 &    0.34 &     0.61 \\
\tabindent Cycleways   &   8.95 &   10.02 &    19.36 \\
\tabindent Residential roads  &  22.88 &   13.50 &    16.57 \\
\tabindent \tabindent No bicycle infrastructure    &  22.61 &   10.22 &    12.74 \\
\tabindent \tabindent W/ bicycle infrastructure    &   0.27 &    3.28 &     1.86 \\
\tabindent Medium roads&  18.37 &   41.06 &    40.93 \\
\tabindent \tabindent No bicycle infrastructure    &  14.34 &    3.07 &     1.90 \\
\tabindent \tabindent W/ bicycle infrastructure    &   4.03 &   37.99 &    39.03 \\
\tabindent Large roads &   2.16 &   21.27 &    18.68 \\
\tabindent \tabindent No bicycle infrastructure    &   1.00 &    3.63 &     0.75 \\
\tabindent \tabindent W/ bicycle lanes      &   0.08 &    1.32 &     1.19 \\
\tabindent \tabindent W/ protected bicycle tracks &   1.08 &   16.32 &    16.74 \\
\textbf{Bicycle route classification}& &  &   \\
\tabindent No classification  &  92.01 &   39.39 &    39.51 \\
\tabindent Cycle superhighway &   2.33 &   23.50 &    23.19 \\
\tabindent Proposed cycle superhighway      &   5.66 &   37.11 &    37.31 \\
\textbf{Land use, cycleways}  & &  &   \\
\tabindent High-rise urban areas   &   0.58 &    4.05 &     5.27 \\
\tabindent Low-rise urban areas    &   3.49 &    1.20 &     1.37 \\
\tabindent Industrial areas   &   0.64 &    1.45 &     2.62 \\
\tabindent Green areas &   1.09 &    2.09 &     4.86 \\
\tabindent Areas near water   &   0.11 &    0.76 &     0.87 \\
\tabindent Open landscape     &   3.04 &    0.48 &     0.69 \\
\textbf{Land use, other infrastructure}     & &  &   \\
\tabindent High-rise urban areas   &   8.25 &   60.07 &    56.60 \\
\tabindent Low-rise urban areas    &  36.54 &    9.47 &    10.25 \\
\tabindent Industrial areas   &   7.92 &    5.59 &     5.55 \\
\tabindent Green areas &  16.05 &    8.87 &     7.41 \\
\tabindent Areas near water   &   1.02 &    4.93 &     3.24 \\
\tabindent Open landscape     &  21.27 &    1.05 &     1.28 \\
\textbf{Elevation gain, $>\,$35 m/km}&   0.16 &    0.01 &     0.02 \\
\textbf{Surface type}  & &  &   \\
\tabindent Asphalt     &  86.25 &   95.85 &    97.62 \\
\tabindent Cobblestones&   0.96 &    2.06 &     1.72 \\
\tabindent Gravel      &  12.79 &    2.09 &     0.65 \\
\textbf{Wrong way}     &   3.54 &    9.15 &     2.17 \\
\bottomrule
\end{tabular}}
 \caption{ Distribution of length shares for various link characteristics. Network indicates the share of the network, observed use indicates the share of the observed trips, and predicted use indicates the share of the predicted flow. }
 \label{tab:usage}
\end{table}

\newpage

\begin{table}[htb] 
 \centering  
\resizebox{\textwidth}{!}{%
\resizebox{\textwidth}{!}{\begin{tabular}{l|rrrr|rrrr|rrrr} \toprule
\textbf{Number of Os and Ds} & \multicolumn{4}{c|}{\textbf{100}} & \multicolumn{4}{c|}{\textbf{200}} & \multicolumn{4}{c}{\textbf{400}} \\  
\textbf{Parameter} & \textbf{Coef.} & \textbf{Std. err.} & \textbf{\textit{P}-val.} &  \textbf{Scaled} & \textbf{Coef.} & \textbf{Std. err.} & \textbf{\textit{P}-val.} &  \textbf{Scaled} & \textbf{Coef.} & \textbf{Std. err.} & \textbf{\textit{P}-val.} &  \textbf{Scaled} \\
\midrule
\textbf{Constant}  &$-\text{0.493}$&$\text{0.035}$&***&$\text{1}$&$-\text{0.456}$&$\text{0.027}$&***&$\text{1}$&$-\text{0.360}$&$\text{0.027}$&***&$\text{1}$\\
\textbf{Infrastructure}  &&&&&&&&&&&&\\
\tabindent Stairs  &$-\text{0.257}$&$\text{0.122}$&*&$\text{0.521}$&$-\text{0.342}$&$\text{0.096}$&***&$\text{0.750}$&$-\text{0.555}$&$\text{0.090}$&***&$\text{1.541}$\\
\tabindent Pedestrian zones  &$-\text{0.141}$&$\text{0.029}$&***&$\text{0.285}$&$-\text{0.121}$&$\text{0.025}$&***&$\text{0.265}$&$-\text{0.106}$&$\text{0.024}$&***&$\text{0.296}$\\
\tabindent Footways  &$-\text{0.073}$&$\text{0.014}$&***&$\text{0.147}$&$-\text{0.090}$&$\text{0.012}$&***&$\text{0.198}$&$-\text{0.099}$&$\text{0.012}$&***&$\text{0.275}$\\
\tabindent Shared paths  &$-\text{0.041}$&$\text{0.019}$&*&$\text{0.082}$&$-\text{0.026}$&$\text{0.016}$&&$\text{0.057}$&$-\text{0.039}$&$\text{0.016}$&*&$\text{0.109}$\\
\tabindent Living streets  &$-\text{0.049}$&$\text{0.054}$&&$\text{0.100}$&$-\text{0.051}$&$\text{0.039}$&&$\text{0.113}$&$-\text{0.076}$&$\text{0.039}$&*&$\text{0.212}$\\
\tabindent Cycleways  &$\text{0.104}$&$\text{0.037}$&**&$-\text{0.211}$&$\text{0.089}$&$\text{0.037}$&*&$-\text{0.195}$&$\text{0.059}$&$\text{0.033}$&&$-\text{0.163}$\\
\tabindent {Residential roads}   &&&&&&&&&&&&\\
\tabindent \tabindent {No bicycle infrastructure}  & ------- & ------- & ------- & ------- & ------- & ------- & ------- & ------- & ------- & ------- & ------- & ------- \\
\tabindent \tabindent W/ bicycle infrastructure  &$\text{0.030}$&$\text{0.023}$&&$-\text{0.062}$&$\text{0.065}$&$\text{0.019}$&***&$-\text{0.142}$&$\text{0.046}$&$\text{0.018}$&*&$-\text{0.128}$\\
\tabindent Medium roads  &$-\text{0.028}$&$\text{0.021}$&&$\text{0.056}$&$-\text{0.005}$&$\text{0.018}$&&$\text{0.010}$&$-\text{0.033}$&$\text{0.018}$&&$\text{0.090}$\\
\tabindent \tabindent {No bicycle infrastructure}  & ------- & ------- & ------- & ------- & ------- & ------- & ------- & ------- & ------- & ------- & ------- & ------- \\
\tabindent \tabindent W/ bicycle infrastructure  &$\text{0.150}$&$\text{0.021}$&***&$-\text{0.304}$&$\text{0.101}$&$\text{0.018}$&***&$-\text{0.221}$&$\text{0.103}$&$\text{0.018}$&***&$-\text{0.285}$\\
\tabindent Large roads  &$-\text{0.035}$&$\text{0.018}$&&$\text{0.070}$&$-\text{0.050}$&$\text{0.015}$&***&$\text{0.110}$&$-\text{0.060}$&$\text{0.015}$&***&$\text{0.167}$\\
\tabindent \tabindent {No bicycle infrastructure}  & ------- & ------- & ------- & ------- & ------- & ------- & ------- & ------- & ------- & ------- & ------- & ------- \\
\tabindent \tabindent W/ painted bicycle lanes  &$-\text{0.023}$&$\text{0.030}$&&$\text{0.046}$&$\text{0.014}$&$\text{0.024}$&&$-\text{0.030}$&$-\text{0.030}$&$\text{0.025}$&&$\text{0.084}$\\
\tabindent \tabindent W/ protected bicycle lanes  &$\text{0.175}$&$\text{0.018}$&***&$-\text{0.355}$&$\text{0.154}$&$\text{0.016}$&***&$-\text{0.338}$&$\text{0.122}$&$\text{0.015}$&***&$-\text{0.339}$\\
\textbf{Bicycle route classification}  &&&&&&&&&&&&\\
\tabindent {No classification}  & ------- & ------- & ------- & ------- & ------- & ------- & ------- & ------- & ------- & ------- & ------- & ------- \\
\tabindent Cycle superhighway  &$\text{0.042}$&$\text{0.016}$&**&$-\text{0.086}$&$\text{0.057}$&$\text{0.014}$&***&$-\text{0.126}$&$\text{0.040}$&$\text{0.014}$&**&$-\text{0.112}$\\
\tabindent Proposed cycle superhighway  &$\text{0.020}$&$\text{0.015}$&&$-\text{0.041}$&$\text{0.055}$&$\text{0.012}$&***&$-\text{0.121}$&$\text{0.038}$&$\text{0.012}$&**&$-\text{0.106}$\\
\textbf{Land use, cycleways} &&&&&&&&&&&&\\
\tabindent High-rise urban areas  &$\text{0.146}$&$\text{0.044}$&***&$-\text{0.296}$&$\text{0.119}$&$\text{0.041}$&**&$-\text{0.260}$&$\text{0.100}$&$\text{0.038}$&**&$-\text{0.277}$\\
\tabindent {Low-rise urban areas}  & ------- & ------- & ------- & ------- & ------- & ------- & ------- & ------- & ------- & ------- & ------- & ------- \\
\tabindent Industrial areas  &$\text{0.278}$&$\text{0.060}$&***&$-\text{0.564}$&$\text{0.221}$&$\text{0.052}$&***&$-\text{0.486}$&$\text{0.183}$&$\text{0.049}$&***&$-\text{0.509}$\\
\tabindent Green areas  &$\text{0.250}$&$\text{0.060}$&***&$-\text{0.506}$&$\text{0.243}$&$\text{0.054}$&***&$-\text{0.534}$&$\text{0.198}$&$\text{0.053}$&***&$-\text{0.549}$\\
\tabindent Areas near water  &$\text{0.105}$&$\text{0.057}$&&$-\text{0.213}$&$\text{0.043}$&$\text{0.052}$&&$-\text{0.093}$&$-\text{0.040}$&$\text{0.048}$&&$\text{0.112}$\\
\tabindent Open landscape  &$-\text{0.131}$&$\text{0.103}$&&$\text{0.266}$&$-\text{0.037}$&$\text{0.123}$&&$\text{0.082}$&$-\text{0.045}$&$\text{0.092}$&&$\text{0.125}$\\
\textbf{Land use, other infrastructure} &&&&&&&&&&&&\\
\tabindent High-rise urban areas  &$\text{0.118}$&$\text{0.031}$&***&$-\text{0.239}$&$\text{0.088}$&$\text{0.024}$&***&$-\text{0.193}$&$\text{0.057}$&$\text{0.025}$&*&$-\text{0.157}$\\
\tabindent {Low-rise urban areas}  & ------- & ------- & ------- & ------- & ------- & ------- & ------- & ------- & ------- & ------- & ------- & ------- \\
\tabindent Industrial areas  &$\text{0.069}$&$\text{0.051}$&&$-\text{0.141}$&$\text{0.072}$&$\text{0.038}$&&$-\text{0.158}$&$\text{0.066}$&$\text{0.039}$&&$-\text{0.184}$\\
\tabindent Green areas  &$\text{0.052}$&$\text{0.046}$&&$-\text{0.106}$&$\text{0.054}$&$\text{0.037}$&&$-\text{0.119}$&$\text{0.041}$&$\text{0.040}$&&$-\text{0.113}$\\
\tabindent Areas near water  &$\text{0.162}$&$\text{0.039}$&***&$-\text{0.329}$&$\text{0.118}$&$\text{0.030}$&***&$-\text{0.259}$&$\text{0.091}$&$\text{0.031}$&**&$-\text{0.251}$\\
\tabindent Open landscape  &$\text{0.159}$&$\text{0.127}$&&$-\text{0.323}$&$\text{0.083}$&$\text{0.096}$&&$-\text{0.182}$&$\text{0.036}$&$\text{0.078}$&&$-\text{0.100}$\\
\textbf{Elevation gain, $>\,$35 m/km}  &$-\text{4.141}$&$\text{6.142}$&&$\text{8.393}$&$-\text{7.559}$&$\text{3.682}$&*&$\text{16.595}$&$-\text{10.552}$&$\text{3.394}$&**&$\text{29.309}$\\
\textbf{Surface type}  &&&&&&&&&&&&\\
\tabindent {Asphalt}  & ------- & ------- & ------- & ------- & ------- & ------- & ------- & ------- & ------- & ------- & ------- & ------- \\
\tabindent Cobblestones  &$-\text{0.019}$&$\text{0.025}$&&$\text{0.038}$&$-\text{0.014}$&$\text{0.019}$&&$\text{0.030}$&$-\text{0.024}$&$\text{0.016}$&&$\text{0.066}$\\
\tabindent Gravel  &$-\text{0.097}$&$\text{0.016}$&***&$\text{0.197}$&$-\text{0.086}$&$\text{0.015}$&***&$\text{0.189}$&$-\text{0.096}$&$\text{0.015}$&***&$\text{0.267}$\\
\textbf{Wrong way}  &$-\text{0.253}$&$\text{0.010}$&***&$\text{0.512}$&$-\text{0.266}$&$\text{0.007}$&***&$\text{0.584}$&$-\text{0.251}$&$\text{0.007}$&***&$\text{0.698}$\\
\midrule
\textbf{\textit{N}} & \multicolumn{4}{c|}{917,525}  & \multicolumn{4}{c|}{1,866,767} & \multicolumn{4}{c}{2,972,994} \\ 
\bottomrule
\end{tabular}}}
\caption{ The estimated parameters for the bicycle route choice model with 100, 200 and 400 Os and Ds. Standard errors are robust. The scaled values are scaled such that the parameter ``Constant'' has a value of 1. Significance levels are *** \textit{P}-value $\leq$ 0.001, ** \textit{P}-value $\leq$ 0.01, and * \textit{P}-value $\leq$ 0.05. \textit{N} denotes the number of rows in the linear regression equation.
 }
 \label{tab:model100200400}
\end{table}

\newpage

\begin{table}[htb] 
 \centering  
\resizebox{\textwidth}{!}{%
\resizebox{\textwidth}{!}{\begin{tabular}{l|rrrr|rrrr|rrrr} \toprule
\textbf{Data} & \multicolumn{4}{c|}{\textbf{Data split A}} & \multicolumn{4}{c|}{\textbf{Full data}} & \multicolumn{4}{c}{\textbf{Data split B}} \\  
\textbf{Parameter} & \textbf{Coef.} & \textbf{Std. err.} & \textbf{\textit{P}-val.} &  \textbf{Scaled} & \textbf{Coef.} & \textbf{Std. err.} & \textbf{\textit{P}-val.} &  \textbf{Scaled} & \textbf{Coef.} & \textbf{Std. err.} & \textbf{\textit{P}-val.} &  \textbf{Scaled} \\
\midrule
\textbf{Constant}  &$-\text{0.474}$&$\text{0.042}$&***&$\text{1}$&$-\text{0.456}$&$\text{0.027}$&***&$\text{1}$&$-\text{0.438}$&$\text{0.032}$&***&$\text{1}$\\
\textbf{Infrastructure}  &&&&&&&&&&&&\\
\tabindent Stairs  &$-\text{0.322}$&$\text{0.145}$&*&$\text{0.678}$&$-\text{0.342}$&$\text{0.096}$&***&$\text{0.750}$&$-\text{0.374}$&$\text{0.120}$&**&$\text{0.853}$\\
\tabindent Pedestrian zones  &$-\text{0.118}$&$\text{0.037}$&**&$\text{0.249}$&$-\text{0.121}$&$\text{0.025}$&***&$\text{0.265}$&$-\text{0.116}$&$\text{0.033}$&***&$\text{0.265}$\\
\tabindent Footways  &$-\text{0.090}$&$\text{0.018}$&***&$\text{0.191}$&$-\text{0.090}$&$\text{0.012}$&***&$\text{0.198}$&$-\text{0.090}$&$\text{0.017}$&***&$\text{0.205}$\\
\tabindent Shared paths  &$-\text{0.033}$&$\text{0.026}$&&$\text{0.070}$&$-\text{0.026}$&$\text{0.016}$&&$\text{0.057}$&$-\text{0.017}$&$\text{0.018}$&&$\text{0.039}$\\
\tabindent Living streets  &$-\text{0.105}$&$\text{0.059}$&&$\text{0.222}$&$-\text{0.051}$&$\text{0.039}$&&$\text{0.113}$&$-\text{0.002}$&$\text{0.053}$&&$\text{0.005}$\\
\tabindent Cycleways  &$\text{0.108}$&$\text{0.054}$&*&$-\text{0.229}$&$\text{0.089}$&$\text{0.037}$&*&$-\text{0.195}$&$\text{0.069}$&$\text{0.053}$&&$-\text{0.158}$\\
\tabindent {Residential roads}   &&&&&&&&&&&&\\
\tabindent \tabindent {No bicycle infrastructure}  & ------- & ------- & ------- & ------- & ------- & ------- & ------- & ------- & ------- & ------- & ------- & ------- \\
\tabindent \tabindent W/ bicycle infrastructure  &$\text{0.036}$&$\text{0.027}$&&$-\text{0.077}$&$\text{0.065}$&$\text{0.019}$&***&$-\text{0.142}$&$\text{0.096}$&$\text{0.025}$&***&$-\text{0.219}$\\
\tabindent Medium roads  &$\text{0.010}$&$\text{0.022}$&&$-\text{0.021}$&$-\text{0.005}$&$\text{0.018}$&&$\text{0.010}$&$-\text{0.019}$&$\text{0.026}$&&$\text{0.044}$\\
\tabindent \tabindent {No bicycle infrastructure}  & ------- & ------- & ------- & ------- & ------- & ------- & ------- & ------- & ------- & ------- & ------- & ------- \\
\tabindent \tabindent W/ bicycle infrastructure  &$\text{0.076}$&$\text{0.022}$&***&$-\text{0.159}$&$\text{0.101}$&$\text{0.018}$&***&$-\text{0.221}$&$\text{0.126}$&$\text{0.027}$&***&$-\text{0.287}$\\
\tabindent Large roads  &$-\text{0.052}$&$\text{0.021}$&*&$\text{0.109}$&$-\text{0.050}$&$\text{0.015}$&***&$\text{0.110}$&$-\text{0.049}$&$\text{0.020}$&*&$\text{0.112}$\\
\tabindent \tabindent {No bicycle infrastructure}  & ------- & ------- & ------- & ------- & ------- & ------- & ------- & ------- & ------- & ------- & ------- & ------- \\
\tabindent \tabindent W/ painted bicycle lanes  &$\text{0.003}$&$\text{0.033}$&&$-\text{0.007}$&$\text{0.014}$&$\text{0.024}$&&$-\text{0.030}$&$\text{0.020}$&$\text{0.034}$&&$-\text{0.045}$\\
\tabindent \tabindent W/ protected bicycle lanes  &$\text{0.161}$&$\text{0.023}$&***&$-\text{0.339}$&$\text{0.154}$&$\text{0.016}$&***&$-\text{0.338}$&$\text{0.145}$&$\text{0.022}$&***&$-\text{0.330}$\\
\textbf{Bicycle route classification}  &&&&&&&&&&&&\\
\tabindent {No classification}  & ------- & ------- & ------- & ------- & ------- & ------- & ------- & ------- & ------- & ------- & ------- & ------- \\
\tabindent Cycle superhighway  &$\text{0.056}$&$\text{0.019}$&**&$-\text{0.119}$&$\text{0.057}$&$\text{0.014}$&***&$-\text{0.126}$&$\text{0.059}$&$\text{0.020}$&**&$-\text{0.135}$\\
\tabindent Proposed cycle superhighway  &$\text{0.025}$&$\text{0.017}$&&$-\text{0.053}$&$\text{0.055}$&$\text{0.012}$&***&$-\text{0.121}$&$\text{0.089}$&$\text{0.016}$&***&$-\text{0.202}$\\
\textbf{Land use, cycleways} &&&&&&&&&&&&\\
\tabindent High-rise urban areas  &$\text{0.133}$&$\text{0.056}$&*&$-\text{0.280}$&$\text{0.119}$&$\text{0.041}$&**&$-\text{0.260}$&$\text{0.104}$&$\text{0.060}$&&$-\text{0.238}$\\
\tabindent {Low-rise urban areas}  & ------- & ------- & ------- & ------- & ------- & ------- & ------- & ------- & ------- & ------- & ------- & ------- \\
\tabindent Industrial areas  &$\text{0.247}$&$\text{0.072}$&***&$-\text{0.520}$&$\text{0.221}$&$\text{0.052}$&***&$-\text{0.486}$&$\text{0.200}$&$\text{0.075}$&**&$-\text{0.458}$\\
\tabindent Green areas  &$\text{0.272}$&$\text{0.081}$&***&$-\text{0.574}$&$\text{0.243}$&$\text{0.054}$&***&$-\text{0.534}$&$\text{0.212}$&$\text{0.073}$&**&$-\text{0.484}$\\
\tabindent Areas near water  &$\text{0.083}$&$\text{0.071}$&&$-\text{0.175}$&$\text{0.043}$&$\text{0.052}$&&$-\text{0.093}$&$\text{0.001}$&$\text{0.075}$&&$-\text{0.003}$\\
\tabindent Open landscape  &$-\text{0.011}$&$\text{0.144}$&&$\text{0.023}$&$-\text{0.037}$&$\text{0.123}$&&$\text{0.082}$&$-\text{0.081}$&$\text{0.212}$&&$\text{0.184}$\\
\textbf{Land use, other infrastructure} &&&&&&&&&&&&\\
\tabindent High-rise urban areas  &$\text{0.121}$&$\text{0.038}$&**&$-\text{0.256}$&$\text{0.088}$&$\text{0.024}$&***&$-\text{0.193}$&$\text{0.056}$&$\text{0.029}$&&$-\text{0.128}$\\
\tabindent {Low-rise urban areas}  & ------- & ------- & ------- & ------- & ------- & ------- & ------- & ------- & ------- & ------- & ------- & ------- \\
\tabindent Industrial areas  &$\text{0.117}$&$\text{0.055}$&*&$-\text{0.246}$&$\text{0.072}$&$\text{0.038}$&&$-\text{0.158}$&$\text{0.032}$&$\text{0.050}$&&$-\text{0.072}$\\
\tabindent Green areas  &$\text{0.093}$&$\text{0.062}$&&$-\text{0.197}$&$\text{0.054}$&$\text{0.037}$&&$-\text{0.119}$&$\text{0.020}$&$\text{0.041}$&&$-\text{0.045}$\\
\tabindent Areas near water  &$\text{0.136}$&$\text{0.048}$&**&$-\text{0.287}$&$\text{0.118}$&$\text{0.030}$&***&$-\text{0.259}$&$\text{0.102}$&$\text{0.037}$&**&$-\text{0.232}$\\
\tabindent Open landscape  &$\text{0.149}$&$\text{0.125}$&&$-\text{0.314}$&$\text{0.083}$&$\text{0.096}$&&$-\text{0.182}$&$-\text{0.018}$&$\text{0.125}$&&$\text{0.040}$\\
\textbf{Elevation gain, $>\,$35 m/km}  &$-\text{9.155}$&$\text{7.159}$&&$\text{19.296}$&$-\text{7.559}$&$\text{3.682}$&*&$\text{16.595}$&$-\text{6.247}$&$\text{3.340}$&&$\text{14.270}$\\
\textbf{Surface type}  &&&&&&&&&&&&\\
\tabindent {Asphalt}  & ------- & ------- & ------- & ------- & ------- & ------- & ------- & ------- & ------- & ------- & ------- & ------- \\
\tabindent Cobblestones  &$-\text{0.015}$&$\text{0.024}$&&$\text{0.032}$&$-\text{0.014}$&$\text{0.019}$&&$\text{0.030}$&$-\text{0.015}$&$\text{0.029}$&&$\text{0.035}$\\
\tabindent Gravel  &$-\text{0.096}$&$\text{0.021}$&***&$\text{0.201}$&$-\text{0.086}$&$\text{0.015}$&***&$\text{0.189}$&$-\text{0.074}$&$\text{0.022}$&***&$\text{0.170}$\\
\textbf{Wrong way}  &$-\text{0.256}$&$\text{0.010}$&***&$\text{0.539}$&$-\text{0.266}$&$\text{0.007}$&***&$\text{0.584}$&$-\text{0.276}$&$\text{0.010}$&***&$\text{0.631}$\\
\midrule

\textbf{\textit{N}} & \multicolumn{4}{c|}{929,569}  & \multicolumn{4}{c|}{1,866,767} & \multicolumn{4}{c}{937,198} \\ 
\bottomrule
\end{tabular}}} 
 \caption{ The estimated parameters for the bicycle route choice model based on data split A, the full dataset, and data split B. Standard errors are robust. The scaled values are scaled such that the parameter ``Constant'' has a value of 1. Significance levels are *** \textit{P}-value $\leq$ 0.001, ** \textit{P}-value $\leq$ 0.01, and * \textit{P}-value $\leq$ 0.05. \textit{N} denotes the number of rows in the linear regression equation.
 }
 \label{tab:modelAFullB}
\end{table}

\newpage

\end{document}